\documentclass[conference]{IEEEtran}
\IEEEoverridecommandlockouts
\usepackage{cite}
\usepackage{amsmath,amssymb,amsfonts}
\usepackage{bm}
\usepackage{algorithm}
\usepackage{algorithmicx}
\usepackage{algpseudocode}
\usepackage{graphicx}
\usepackage{textcomp}
\usepackage{xcolor}
\usepackage{multirow}
\usepackage{caption}
\usepackage{booktabs}
\usepackage{caption}
\usepackage{subfigure} 
\usepackage{threeparttable} 
\usepackage{amsthm}
\newtheorem{definition}{Definition}

\usepackage[switch]{lineno} 
\def\BibTeX{{\rm B\kern-.05em{\sc i\kern-.025em b}\kern-.08em
    T\kern-.1667em\lower.7ex\hbox{E}\kern-.125emX}}
\begin{document}
\title{Neural Node Matching for Multi-Target Cross Domain Recommendation}
\newcommand{\wujiang}[1]{{\textcolor[rgb]{0,0,0}{#1}}}

\author{
    \IEEEauthorblockN{Wujiang Xu $^{ \dag *}$, Shaoshuai Li $^{ \dag *}$, Mingming Ha $^{ \dag\xi}$, Xiaobo Guo $^{ \ddag\S}$, Qiongxu Ma $^{ \dag}$,\\ Xiaolei Liu $^{ \dag}$, Linxun Chen $^{ \dag}$, Zhenfeng Zhu $^{ \ddag\S}$ \thanks{ * Joint first author.  \; \S\;Correspondding author.}}
    \IEEEauthorblockA{$^\dag$ MYbank, Ant Group, Hangzhou, China}
    \IEEEauthorblockA{$^\ddag$ Institute of Information Science, Beijing Jiaotong University, Beijing, China}
    \IEEEauthorblockA{$^\xi$ Automation and Electrical Engineering, University of Science and Technology Beijing, Beijing, China}
    \IEEEauthorblockA{\{xuwujiang.xwj,lishaoshuai.lss,hamingming.hmm,qiongxu.mqx,liuxiaolei.lxl,\\ linxun.clx\}@mybank.cn, \{xb\_guo,zhfzhu\}@bjtu.edu.cn}
}


\maketitle
\thispagestyle{plain}
\pagestyle{plain}
\begin{abstract}
Multi-Target Cross Domain Recommendation(CDR) has attracted a surge of interest recently, \textcolor{black}{which intends to improve the recommendation performance in multiple domains (or systems) simultaneously. Most existing multi-target CDR frameworks primarily rely on the existence of the majority of overlapped users across domains}. However, \textcolor{black}{general practical CDR scenarios cannot meet the strictly overlapping requirements and only share a small margin of common users across domains}. Additionally, \textcolor{black}{the majority of users have quite a few historical behaviors in such small-overlapping CDR scenarios}. To tackle the aforementioned issues, we propose a simple-yet-effective neural node matching based framework for more general CDR settings, i.e., only (few) partially overlapped users exist across domains and \textcolor{black}{most overlapped as well as non-overlapped users do have sparse interactions}. \textcolor{black}{The present framework} mainly contains two modules: (i) intra-to-inter node matching module, and (ii) intra node complementing module. Concretely, the first module conducts intra-knowledge fusion within each domain and subsequent inter-knowledge fusion across domains by fully connected user-user homogeneous graph information aggregating.
By doing this, the knowledge of all users, especially the non-overlapping users, could be well extracted and transferred without relying heavily on overlapping users. The second module introduces user-item matching to complement the potential missing interactions for each user and correct his/her under-represented representations, especially for the users with observed sparse interactions. Essentially, \textcolor{black}{companion objectives are also inserted into each module to guide the knowledge transferring procedures, which leads to positive effects on multiple domains simultaneously.} Extensive experiments on four multi-target CDR tasks from both public and real-world large-scale financial industry datasets demonstrate the remarkable performance of our proposed approach. \wujiang{Our code is publicly available at the link: https://github.com/WujiangXu/NMCDRR.}

\end{abstract}


\section{Introduction}
\textcolor{black}{With the rapid development of the digital era, an increasing number of users participate in multiple domains (platforms) for various purposes. Since the overlapped users across domains are likely to have similar interests, it is possible to boost the recommendation performance of other (target) domains by using information collected from several (source) domains, which is the core idea of Cross-Domain Recommendation (CDR).} \textcolor{black}{According to different recommendation scenarios, CDR problems can be generally classified into two categories: single-target CDR and multi-target CDR.} \textcolor{black}{The conventional single-target CDR aims at using source domain information to enhance recommendation performance in target domain. Multi-target CDR expects to improve the recommendation performance in multiple domains simultaneously and has recently attracted increasing attention. To achieve a valid multi-target CDR performance, several excellent works focusing on feature combination \cite{zhu2020deep,zhang2018cross,zhao2019cross} or bi-directional transfer mapping strategies \cite{hu2018conet,zhu2020graphical,ouyang2020minet,salah2021towards} have been proposed. However, these learning frameworks primarily assume the existence of fully overlapped users across domains, which is difficult to cope with general partially overlapped CDR scenarios. In this work, we focus on developing an effective multi-target CDR model for the more general CDR settings with (few) partially overlapped users}. This intention faces two critical challenges. 


\textbf{CH1.} For multiple domains with only (few) partially overlapped users, how to improve the recommendation performance for multi-target CDR tasks? 

\textcolor{black}{Most previous multi-target CDR methods cannot be directly extended to partially overlapped CDR settings, especially for few overlapped users across domains.} To explore the knowledge of the non-overlapped users, several recent efforts \cite{zhu2019dtcdr, li2020ddtcdr,li2021dual, cui2020herograph} try to introduce graphic deep learning to get both overlapped and non-overlapped user embeddings by collecting user-item interactions. However, such graph-based CDR approaches still rely heavily on overlapped users (more than 80$\%$ are common users across domains) to bridge connections among multiple domains and then conduct knowledge aggregation and transition processes to get the representative embeddings of non-overlapped users. \textcolor{black}{Nevertheless, in small-overlapping CDR scenarios, such above methods have great limitations.} Therefore, it is challenging to guarantee the multi-target cross-domain recommendation performance with only quite a few overlapping users. To mitigate the small overlapping problem, the recent model PTUPCDR \cite{zhu2022personalized} proposes a meta network fed with pre-trained user/item representations to generate personalized bridge functions to transfer preference for each user, while VDEA \cite{liu2022exploiting} utilizes VAE framework to exploit user domain-invariant embedding across different domains. However, such methods treat all users equally and do not pay special attention to the majority of data-sparse users (i.e., users with quite a few historical behaviors), resulting in inferior knowledge fusing and transferring effectiveness, which leads to the second challenge.
\begin{figure}[t!]
\centering{
\includegraphics[width=0.35\textwidth]{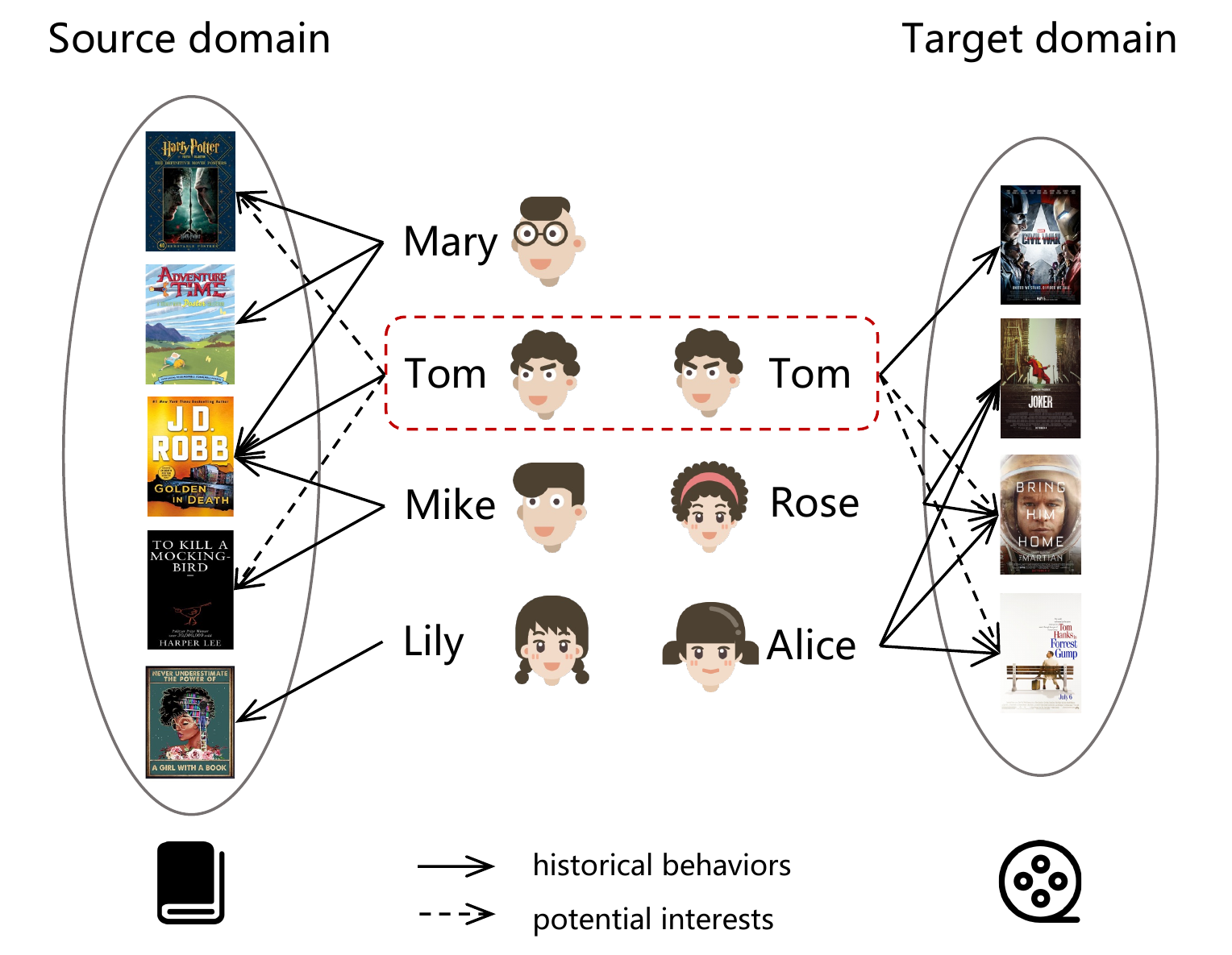}}
\captionsetup{font={footnotesize}}
\caption{The partially overlapped CDR scenarios.}
\label{intro}
\end{figure}

\textbf{CH2.} For the majority of overlapping as well as non-overlapping users with few historical behaviors, how to improve the recommendation performance of multiple domains simultaneously?

The user-item interactions of the most real-world recommendation systems generally present the intrinsic long-tailed distribution, which means that a majority of users (i.e., tail users) have very few interactions and a few users (i.e., head users) have a huge number of interactions.
Consider the toy example shown in Fig. \ref{intro}, Mary and Alice could be roughly treated as head users, while Mike, Lily, Rose, and Tom could be treated as tail users. Essentially, both the overlapping and non-overlapping tail users may be under-represented based on their observed sparse interactions, since most representing CDR models (e.g., Herograph \cite{cui2020herograph}) are easily dominated by the data-rich users. As shown in Fig. \ref{intro}, Tom, Mary, and Mike all like reading romantic book \textit{JO ROBB}, but Mary and Mike still like reading magic book \textit{Harry Potter} and educational book \textit{To Kill a Mockingbird} respectively, thus Tom may also have potential interests in \textit{Harry Potter} and \textit{To Kill a Mockingbird}. Consequently, based on the only sparse interaction with \textit{JO ROBB}, a biased representation of Tom would be got and used to conduct ranking recommendation tasks, which may lead to inferior performance. \textcolor{black}{How to get informative embeddings of tail users by complementing their potential missing interactions becomes another practical challenge, which is frequently ignored in existing multi-target CDR models.}

\begin{figure*}[h!]\color{blue}
\centering{
\includegraphics[width=0.9\textwidth]{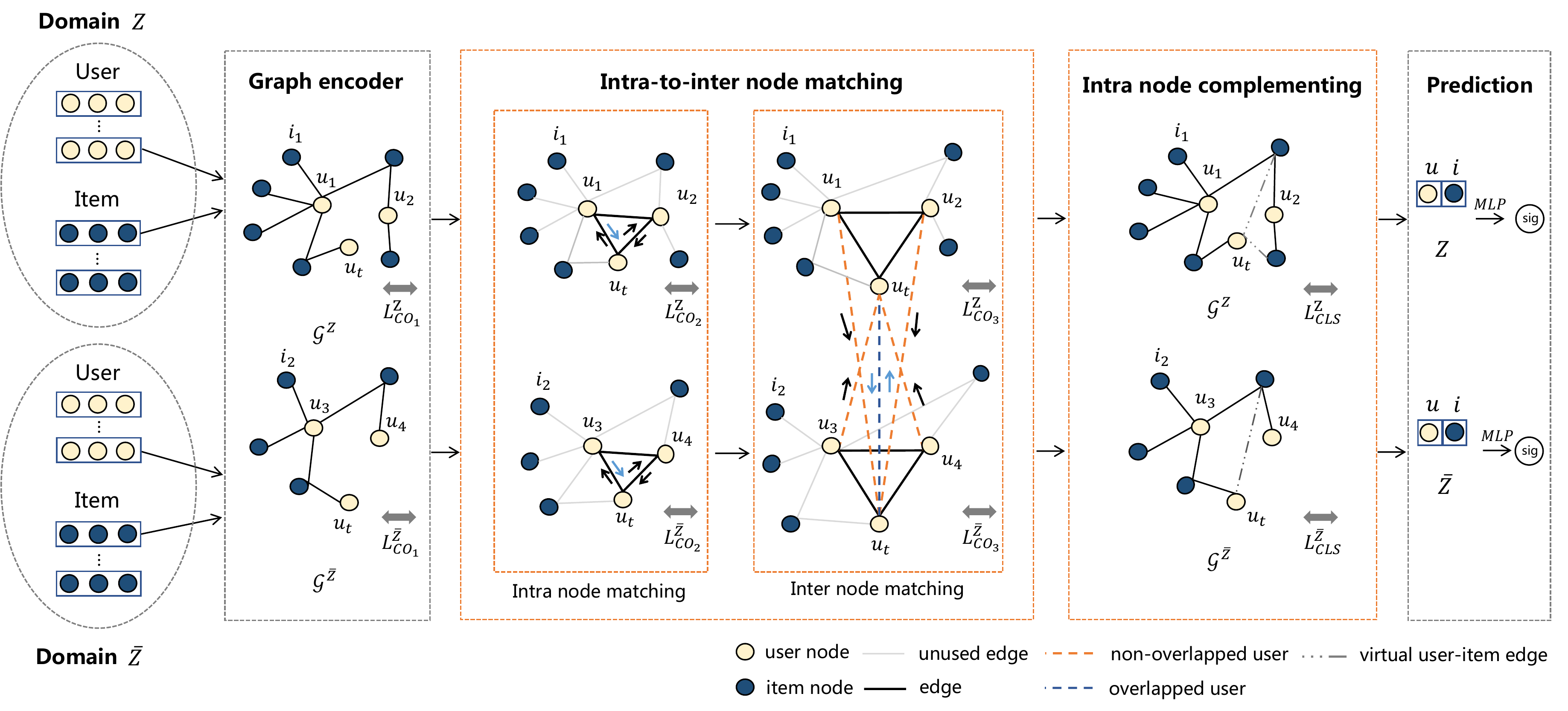}}
\captionsetup{font={footnotesize}}
\caption{\textcolor{black}{Overview of NMCDR. In the intra node matching component, the black (i.e., $u_t$ $\rightarrow$ $u_1$) and blue (i.e., $u_1$ $\rightarrow$ $u_t$) solid arrow denote two different types of messages propagated by tail and head users respectively. In the inter node matching component, the orange (i.e., $u_1$ $\rightarrow$ $u_t$) and blue (i.e., $u_t$ $\leftrightarrow$ $u_t$) dashed line denotes the knowledge fusing bridge for non-overlapping and overlapping users respectively. The black dashed line in intra node complementing component denotes the predicted virtual user-item interactions. $L_{CO}^Z$, $L_{CO}^{\bar{Z}}$ in each component represent companion objective loss, while $L_{CLS}^Z$, $L_{CLS}^{\bar{Z}}$ indicates the final prediction loss.}}
\label{Fig_framework}
\end{figure*}

\textcolor{black}{\textbf{Our Approach.} To address the aforementioned challenges, we propose a novel neural \textbf{n}ode \textbf{m}atching based framework for multi-target \textbf{CDR} with only partially overlapped users, named as \textbf{NMCDR}. Our model mainly contains two modules, namely \textit{intra-to-inter node matching module and intra node complementing module}, which corresponds to tackle \textbf{CH1} and \textbf{CH2}, respectively. The intra-to-inter node matching module further contains two components, e.g., intra node matching component and inter node matching component, as shown in Fig. \ref{Fig_framework}. In detail, \textbf{To tackle \textbf{CH1}}, a heterogeneous graph encoder is used to model the direct user-item interaction. Then, inspired by \cite{alon2020bottleneck}, the intra node matching component designs a fully connected user-user homogeneous graph within every single domain and conducts user-to-user knowledge fusion. With this operation, the knowledge flow within each domain can be eased and each user can directly consider nodes beyond their original neighbors. The enhanced user representations are then fed into the inter node matching component, which conducts user-to-user knowledge fusion for both overlapping and non-overlapping users. By doing this, the knowledge of all users, especially the non-overlapping users, could be well extracted and transferred without relying heavily on overlapping users. \textbf{To tackle \textbf{CH2}}, the intra node complementing module shown in Fig. \ref{Fig_framework} conducts user-to-item matching and tends to correct the biased representations by complementing the potential missing interactions for each user, especially for the tail users. Moreover, we insert companion objectives into each module to guide knowledge fusion procedures and guarantee the simultaneous performance improvement of multiple domains.}

\textbf{Contributions.} Overall, our major contributions can be summarized as follows:

(1) \textcolor{black}{We develop a novel neural node matching based framework to address the multi-target CDR scenarios with (few) partially overlapped users, which employ intra-to-inter node matching module and intra node complementing module to efficiently and effectively lead positive recommendation effect on all domains.}

(2) To obtain representative user embeddings, especially the tail users with observed sparse interactions, we consider complementing the potential missing information for each user to correct the biased representation for ranking recommendation tasks. To our knowledge, this paper is the first work to correct the potential interaction bias in multi-target CDR scenarios. 

(3) We conduct extensive experiments on four CDR scenarios including both public and real-world large-scale financial industry datasets to show the remarkable performance of the proposed approach in kinds of evaluation metrics. Besides, we provide theoretical insight to evaluate our model stability.

\section{Methodology}

\subsection{Problem Formulation}
In this work, we consider a general partially overlapped multi-target CDR scenario composed of two domains $Z$ and $\bar{Z}$. Let $\mathcal{G}^Z=(\mathcal{U}^Z, \mathcal{V}^Z, \mathcal{E}^Z)$ and $\mathcal{G}^{\bar{Z}}=(\mathcal{U}^{\bar{Z}}, \mathcal{V}^{\bar{Z}}, \mathcal{E}^{\bar{Z}})$ be the domain data, where $\mathcal{U}$, $\mathcal{V}$, $\mathcal{E}$ are the user set, item set and edge set for each domain. Particularly, the overlapped user subset is defined as  $\mathcal{U}^{O}=\mathcal{U}^{Z} \cap \mathcal{U}^{\bar{Z}}$, while the non-overlapped user subset for each domain is $\mathcal{U}^Z_{non} = \mathcal{U}^Z\backslash \mathcal{U}^{O}$ and $\mathcal{U}^{\bar{Z}}_{non} = \mathcal{U}^{\bar{Z}}\backslash \mathcal{U}^{O}$ respectively. Given the observed data, multi-target CDR aims to improve the recommendation performance of both domains simultaneously by fusing and transferring knowledge across domains.



\subsection{Overview}
Fig. \ref{Fig_framework} illustrates the pipeline of our proposed neural node matching based framework for multi-target cross-domain recommendation (NMCDR), which mainly consists of two modules: intra-to-inter node matching module and intra node complementing module.
We first utilize a graph encoder to model direct user-item interactions by building a heterogeneous user-item graph. Then, the learned user representation will be fed into the intra-to-inter node matching module, which further consists of an intra node matching component and an inter node matching component. The intra node matching component enhances user representation by conducting within-domain fully connected user-to-user information aggregating, while the inter node matching component tends to transfer knowledge for both overlapping and non-overlapping users by conducting cross-domain fully connected user-to-user information aggregating. After that, the intra node complementing module corrects the biased user representation by exploiting the potential missing user-item interactions for each user, especially for the tail users.
Finally, the prediction layer outputs the affinity score of a user-item pair for each domain individually.

\subsection{Heterogeneous Graph Encoder}
For domain $Z$, we first construct a heterogeneous user-item graph $\mathcal{G}^Z$ to learn the users' preference for each domain explicitly. For each user and each item in $\mathcal{G}^Z$, we introduce corresponding embedding vectors $\bm{u} \in  \mathbb{R}^{D}$ and $\bm{v} \in  \mathbb{R}^{D}$ as their representations, where $D$ denotes the embedding dimension. Specifically, the initialized representations for $N$ users and $M$ items of the domain $Z$ can be obtained from the following look-up table:
\begin{small}
\begin{equation}
\bm{E}^{Z} = \Big [ \bm{u}_{1}^{Z},\cdots,\bm{u}_{N}^{Z} \;,\;  \bm{v}_{1}^{Z},\cdots,\bm{v}_{M}^{Z}   \Big ].
\label{embl1}
\end{equation}
\end{small}
Formally,  $U^Z \in \mathbb{R}^{|  \mathcal{U}^{Z}  | \times D}$ and  $V^{Z} \in \mathbb{R}^{|  \mathcal{V}^{Z}  | \times D}$ are the learnable embeddings for the user/item sets $\mathcal{U}$ and $\mathcal{V}$.
To encoder the explicit user-item interactions within domain $Z$, we utilize a vanilla GNN operation, which can be formulated as message construction and message aggregation procedures.

\noindent \textbf{Message Construction.} Given a user-item pair ($u_{i}^{Z}$, $v_{j}^{Z}$, $e_{u_{i}v_{j}}^{Z}$), we define the message from the item $v_{j}^{Z}$ to user $u_{i}^{Z}$ with edge $e_{u_{i}v_{j}}^{Z}$ as:
\vspace{-15pt}

\begin{equation}
\begin{small}
\bm{m}_{u_{i}^{Z} \leftarrow v_{j}^{Z}} = f_{u_i}(\bm{v}_{j}^{Z},e_{u_{i}v_{j}}^{Z}),
\label{uimc_Z}
\end{small}
\end{equation}
where $\bm{m}_{u_{i}^{Z} \leftarrow v_{j}^{Z}}$ is the transferred message representation. $f_{u_i}(\cdot)$ denotes the message mapping function in the user-item graph, which takes item embedding and edge embedding as the input. In practice, we instantiate $f_{u_i}(\cdot)$ and rewrite Eq. \ref{uimc_Z} as follows:
\begin{small}
\begin{equation}
\bm{m}_{u_{i}^{Z} \leftarrow v_{j}^{Z}} =  \frac{1}{| \mathcal{N}_{u_{i}^{Z}} |} (\bm{v}_{j}^{Z}\bm{W}_{hge}^{Z}+\bm{b}_{hge}^{Z})e_{u_{i}v_{j}}^{Z},
\label{m1funcX}
\end{equation}
\end{small}
where $\bm{W}_{hge}^{Z} \in \mathbb{R}^{D\times D_{hge}}$ and $\bm{b}_{hge}^{Z}  \in \mathbb{R}^{D_{hge}}$ are the trainable weight matrix and bias vector during information propagation. If user $u_{i}^{Z}$ interacts with item $v_{j}^{Z}$, then $e_{u_{i}v_{j}}^{Z}$ is set to 1, else 0. $D_{hge}$ is the transformation dimension. Following the graph neural networks operation \cite{wang2019neural,kipf2016semi}, we set $1/| \mathcal{N}_{u_{i}^{Z}}|$ as the graph Laplacian norm, where $\mathcal{N}_{u_{i}^{Z}}$ denotes the first-hop neighbors of user $u_{i}^{Z}$. It is worth noticing that the message mapping function can be replaced with any proposed graph neural network kernels such as GCN \cite{kipf2016semi} and GAT \cite{velivckovic2017graph}.

\noindent \textbf{Message Aggregation.} In this stage, we aggregate all the messages from the user's neighborhood to obtain his/her representation. The aggregation function is formulated as:
\begin{small}
\begin{equation}
\bm{u}_{g1_{i}}^{Z} = \mathrm{ReLU}(\tilde{\bm{m}}_{u_{i}^{Z}} + \;\sum\limits_{v_{j}\in \mathcal{N}_{u_{i}^{Z}}} \bm{m}_{u_{i}^{Z} \leftarrow v_{j}^{Z}}),
\label{m1aggrX}
\end{equation}
\end{small}
where $\bm{u}_{g1_{i}}^{Z}$ denotes the representation vector of $u_{i}^{Z}$ after heterogeneous graph encoder, which consists of a self-mapping message $\tilde{\bm{m}}_{u_{i}^{Z}}=\bm{u}_{i}^{Z}\bm{W}_{hge}^{Z}$ and aggregated neighboring message. ReLU is the activation function. 

\textbf{For domain $\bar{Z}$, to obtain the user representations, we construct the heterogeneous user-item graph $\mathcal{G}^{\bar{Z}}$ and conduct subsequent message construction as well as message aggregation operations being similar with domain $Z$.}

\subsection{Intra-to-Inter node Matching Module}
As shown in Fig. \ref{Fig_framework}, the user representation obtained by the heterogeneous graph encoder is then fed to the intra-to-inter node matching module, which intends to effectively fuse and transfer the knowledge of both overlapping and non-overlapping users across domains without relying heavily on common users. The intra-to-inter node matching module further contains two components, e.g., intra node matching component and inter node matching component. In what follows, we will illustrate each component in detail.


\subsubsection{Intra Node Matching Component}
In most previous GNN-based multi-target CDR methods \cite{zhu2019dtcdr, li2020ddtcdr,li2021dual, cui2020herograph}, the user representations derived from the graph encoder within each domain are directly utilized for cross domain knowledge transferring. \textcolor{black}{However, as the majority of users of each domain do have few historical interactions, these tail users may be under-represented based on their observed sparse interactions and impair the subsequent cross domain knowledge transferring. Thus we argue that it is critical to priory perform intra-domain knowledge fusion for each domain.} Inspired by \cite{alon2020bottleneck}, we design a simple but effective fully connected homogeneous user-user graph and conduct direct user-to-user information aggregating like node-level matching in graph matching procedure \cite{caetano2009learning}, which could enable each user to interact directly and consider nodes beyond their original neighbors and thus ease the knowledge flow within each domain.

\noindent \textbf{Message Construction.} \textcolor{black}{During intra domain knowledge fusing, we believe that information bridges between head users and tail users should also be varied. Thus, for domain $Z$, we first distinguish a user $u_i^Z$ as \textit{head} user or \textit{tail} user as follows:}
\begin{small}
\begin{equation}
u_{i}^{Z}=\left\{
\begin{array}{rcl}
\textit{head}\;user\;, & & | {\mathcal{N}_{u_{i}^{Z}} |\leq \mathcal{K}_{head}}\\
\textit{tail}\;user \;, & & |{\mathcal{N}_{u_{i}^{Z}} | > \mathcal{K}_{head}}
\end{array} \right.
\label{head_tailX}
\end{equation}
\end{small}
where $\mathcal{K}_{head}$ denotes the head/tail user discrimination threshold. $|\mathcal{N}_{u_i}^{Z}|$ represents the number of the items interacted by each user $u_i^Z$ in the domain $Z$.

Then, by constructing a fully connected user-user homogeneous graph $\mathcal{G}_{intra}^Z$, for each user $u_{i}^{Z}$, the matching message from a head user $u_{k}^{Z}$ and a tail user $u_{l}^{Z}$ are formulated as:  
\begin{small}
\begin{equation}
\bm{m}^{head}_{u_{i}^{Z} \leftarrow u_{k}^{Z}} = f_{head}(\bm{u}_{g1_{i}}^{Z},\bm{u}_{g1_{k}}^{Z})
\label{uheadm_X}
\end{equation}
\end{small}
\begin{small}
\begin{equation}
\bm{m}^{tail}_{u_{i}^{Z} \leftarrow u_{l}^{Z}} = f_{tail}(\bm{u}_{g1_{i}}^{Z},\bm{u}_{g1_{l}}^{Z})
\label{utailm_X}
\end{equation}
\end{small}
\wujiang{where $\bm{m}^{head}_{u_{i}^{Z} \leftarrow u_{k}^{Z}}$, $\bm{m}^{tail}_{u_{i}^{Z} \leftarrow u_{l}^{Z}}$ are the message representations from head and tail user respectively. $f_{head}(\cdot)$ and $f_{tail}(\cdot)$ represent corresponding message mapping function in graph $\mathcal{G}_{intra}^Z$.
Besides, $\bm{m}^{head}_{u_{i}^{Z} \leftarrow u_{k}^{Z}}$ and $\bm{m}^{tail}_{u_{i}^{Z} \leftarrow u_{l}^{Z}}$ are represented by the blue and black arrow in \textit{Intra node matching} module of Fig. \ref{Fig_framework}, respectively.
In practice, we implement $f_{head}(\cdot)$ and $f_{tail}(\cdot)$ as:}
\begin{small}
\begin{equation}
\begin{array}{c}
\bm{m}^{head}_{u_{i}^{Z} \leftarrow u_{k}^{Z}} =  \frac{1}{| \mathcal{N}_{u_{i}^{Z}}^{head} |} (\bm{u}_{g1_{k}}^{Z}\bm{W}_{head}^{Z}+\bm{b}_{head}^{Z}), \vspace{1.2ex} \\
\bm{m}^{tail}_{u_{i}^{Z} \leftarrow u_{l}^{Z}} = \frac{1}{| \mathcal{N}_{u_{i}^{Z}}^{tail} |} (\bm{u}_{g1_{l}}^{Z}\bm{W}_{tail}^{Z}+\bm{b}_{tail}^{Z}).
\label{m2}
\end{array}
\end{equation}
\end{small}
where $\bm{W}_{head}^{Z}$ $\in$ $\mathbb{R}^{D_{hge}\times D_{igm}}$,  $\bm{W}_{tail}^{Z}$ $\in$ $\mathbb{R}^{D_{hge}\times D_{igm}}$, and $\bm{b}_{head}^{Z}$ $\in \mathbb{R}^{D_{igm}}$, $\bm{b}_{tail}^{Z}$ $\in \mathbb{R}^{D_{igm}}$ are the trainable weight matrices and bias vectors to transfer information from head and the tail users respectively. $D_{igm}$ is a customized transformation size. Similarly as above, we set $1/| \mathcal{N}_{u_{i}^{Z}}^{head}|$ and $1/| \mathcal{N}_{u_{i}^{Z}}^{tail}|$  as the graph Laplacian norm, where $\mathcal{N}_{u_{i}^{Z}}^{head}$ and $\mathcal{N}_{u_{i}^{Z}}^{tail}$ denotes the fully-connected head and tail user set for $u_{i}^{Z}$. 

\noindent \textbf{Message Aggregation.}
For user $u_{i}^{Z}$, the message extracted from the head and tail users are first aggregated according to: 
\begin{small}
\begin{equation}
\begin{array}{c}
\bm{u}_{head_{i}}^{Z} = \mathrm{ReLU}(\sum\limits_{u_{k}\in \mathcal{N}_{u_{i}^{Z}}^{head}} \bm{m}^{head}_{u_{i}^{Z} \leftarrow u_{k}^{Z}}), \vspace{0.8ex}\\
\bm{u}_{tail_{i}}^{Z} = \mathrm{ReLU}(\sum\limits_{u_{l}\in \mathcal{N}_{u_{i}^{Z}}^{tail}} \bm{m}^{tail}_{u_{i}^{Z} \leftarrow u_{l}^{Z}}).
\label{m1aggrX}
\end{array}
\end{equation}
\end{small}
\vspace{-12pt}

Then, instead of direct concatenating or adding operation, we design a fine-grained gating mechanism to fuse these two kinds of messages as follows:
\vspace{-12pt}

\begin{small}
\begin{align}
\bm{H}_{igm}^{Z} =&\; \sigma(\bm{u}_{head_{i}}^{Z}\bm{W}_{h}^{Z}+ \bm{b}_{h}^{Z}+ \bm{u}_{tail_{i}}^{Z}\bm{W}_{t}^{Z}+ \bm{b}_{t}^{Z}),
\vspace{1.5ex}\nonumber\\
\bm{u}_{g2_{i}^{'}}^{Z} =&\; \tanh{((1-\bm{H}_{igm}^{Z}) \odot \bm{u}_{head_{i}}^{Z}+\bm{H}_{igm}^{Z} \odot \bm{u}_{tail_{i}}^{Z})}.
\label{gate}
\end{align}
\end{small}
where $\bm{u}_{g2_{i}^{'}}^{Z}$ denotes the fused message vector for user $u_{i}^{Z}$. $\sigma(\cdot)$ is the sigmoid function and  $\odot$ is the Hadamard product. $\bm{W}_{h}^{Z}$ $\in \mathbb{R}^{D_{igm}\times D_{igm}}$, $\bm{W}_{t}^{Z} \in \mathbb{R}^{D_{igm}\times D_{igm}}$,  and $\bm{b}_{h}^{Z}$ $\in \mathbb{R}^{D_{igm}}$, $\bm{b}_{t}^{Z}$ $\in \mathbb{R}^{D_{igm}}$ are the trainable parameters. The representation of $u_{i}^{Z}$ after intra node matching component is obtained as:
\begin{small}
\begin{equation}
\begin{array}{cc}
\bm{u}_{g2_{i}}^{Z} = \bm{u}_{g2_{i}^{'}}^{Z} + \bm{u}_{g1_{i}}^{Z}.
\label{add_intra}
\end{array}
\end{equation}
\end{small}

\textbf{For domain $\bar{Z}$, we construct the fully connected homogeneous user-user graph $\mathcal{G}_{intra}^{\bar{Z}}$ and conduct subsequent intra knowledge fusion being similar to domain $Z$.}

\subsubsection{Inter Node Matching Component}
\textcolor{black}{In this component, we conduct the node matching operation for both overlapped and non-overlapped users to fuse and transfer knowledge across domains. Being similar to intra node matching component, we introduce a fully connected cross-domain user-user graph and treat overlapped users and non-overlapped users with different message transferring bridges.}


\noindent \textbf{Message Construction.} 
The fully connected cross-domain user-user homogeneous graph $\mathcal{G}_{inter}$ indicates that each user in one domain is fully connected to the users in the other domain. For each user $u_{i}^{Z}$, given the overlapped and non-overlapped user-user pairs ($u_{i}^{Z}$,$u_{i}^{\bar{Z}}$) and ($u_{i}^{Z}$,$u_{r}^{\bar{Z}}$), the cross-domain message transferring is formulated as: 
\begin{small}
\begin{equation}
\begin{array}{c}
\bm{m}^{self}_{u_{i}^{Z} \leftarrow u_{i}^{\bar{Z}}} = f_{self}(\bm{u}_{g2_{i}}^{Z},\bm{u}_{g2_{i}}^{\bar{Z}}), \vspace{1.2ex} \\
\bm{m}^{other}_{u_{i}^{Z} \leftarrow u_{r}^{\bar{Z}}} = f_{other}(\bm{u}_{g2_{i}}^{Z},\bm{u}_{g2_{k}}^{\bar{Z}}), \vspace{1.2ex}
\label{ucm_Y}
\end{array}
\end{equation}
\end{small}
where $\bm{m}^{self}_{u_{i}^{Z} \leftarrow u_{i}^{\bar{Z}}}$ denotes the cross-domain message representation from the same (overlapped) user in domain $\bar{Z}$ while $\bm{m}^{other}_{u_{i}^{Z} \leftarrow u_{r}^{\bar{Z}}}$ is the cross-domain message representation from other (non-overlapped) users in domain $\bar{Z}$. $f_{self}$ and $f_{other}$ are message mapping functions and we instantiate them as:
\vspace{-6pt}

\begin{small}
\begin{equation}
\begin{array}{c}
\bm{m}^{self}_{u_{i}^{Z} \leftarrow u_{i}^{\bar{Z}}} =  \bm{u}_{g2_{i}}^{\bar{Z}}\bm{W}_{self}^{Z}+\bm{b}_{self}^{Z}, \vspace{1.2ex} \\

\bm{m}^{other}_{u_{i}^{Z} \leftarrow u_{r}^{\bar{Z}}} = \frac{1}{| \mathcal{N}_{u_{i}^{Z}}^{cdr} |} (\bm{u}_{g2_{k}}^{\bar{Z}}\bm{W}_{other}^{Z}+\bm{b}_{other}^{Z}), \vspace{1.2ex} 
\label{m3}
\end{array}
\end{equation}
\end{small}
\vspace{-4pt}

\wujiang{where $\bm{W}_{self}^{Z} \in$ $\mathbb{R}^{D_{igm}\times D_{cgm}}$,  $\bm{W}_{other}^{Z} \in$ $\mathbb{R}^{D_{igm}\times D_{cgm}}$ and $\bm{b}_{self}^{Z} \in \mathbb{R}^{D_{cgm}}$, $\bm{b}_{other}^{Z} \in \mathbb{R}^{D_{cgm}}$ are the trainable parameters to transfer the cross-domain knowledge among the overlapped and non-overlapped users. $D_{cgm}$ is the transformation dimension. We set $1/| \mathcal{N}_{u_{i}^{Z}}^{cdr}|$ as the graph Laplacian norm, where $\mathcal{N}_{u_{i}^{Z}}^{cdr}$ denotes the number of the fully connected non-overlapped users from other domain $\bar{Z}$ with respect to $u_{i}^{Z}$. As shown in \textit{Inter node matching} module of Fig. \ref{Fig_framework}, $\bm{m}^{self}_{u_{i}^{Z} \leftarrow u_{i}^{\bar{Z}}}$ and $\bm{m}^{other}_{u_{i}^{Z} \leftarrow u_{r}^{\bar{Z}}}$ are represented by the blue and the black solid arrow, respectively.}

\noindent \textbf{Message Aggregation.} For user $u_{i}^{Z}$, the aggregated message representations from the overlapped and the non-overlapped users are computed as follows:
\vspace{-15pt}

\begin{small}
\begin{align}
\bm{u}_{self_{i}}^{Z} =& \mathrm{ReLU}(\bm{m}^{self}_{u_{i}^{Z} \leftarrow u_{i}^{\bar{Z}}}), \vspace{0.8ex}\nonumber\\
\bm{u}_{other_{i}}^{Z} =& \mathrm{ReLU}(\sum\limits_{u_{r}\in \mathcal{N}_{u_{i}^{Z}}^{cdr}} \bm{m}^{other}_{u_{i}^{Z} \leftarrow u_{r}^{\bar{Z}}}).
\label{m3aggrX}
\end{align}
\end{small}
Then, we fuse the user representation $\bm{u}_{g2_{i}}$ with the overlapped cross-domain information $\bm{u}_{self_{i}}$ as follows:
\begin{small}
\begin{equation}
\begin{array}{c}
\bm{u}_{g3_{i}^{*}}^{Z} = \bm{u}_{g2_{i}}^{Z}\bm{W}_{cross}^{Z}+\bm{u}_{self_{i}}^{Z}(1-\bm{W}_{cross}^{\bar{Z}}), \vspace{0.8ex}\\
\bm{u}_{g3_{i}^{*}}^{\bar{Z}} = \bm{u}_{g2_{i}}^{\bar{Z}}\bm{W}_{cross}^{\bar{Z}}+\bm{u}_{self_{i}}^{\bar{Z}}(1-\bm{W}_{cross}^{Z}),
\label{m3selffuse}
\end{array}
\end{equation}
\end{small}
\vspace{-6pt}

where $\bm{W}_{cross}^Z$ $\in$ $\mathbb{R}^{D_{cgm}\times D_{cgm}}$ and $\bm{W}_{cross}^{\bar{Z}}$ $\in$ $\mathbb{R}^{D_{cgm}\times D_{cgm}}$ denote the transformation matrices. 
Then, we utilize a gating network to further enhance the user representation by adopting the cross-domain message from the non-overlapped users. Mathematically, the gating operation denotes as:

\begin{small}
\begin{align}
\bm{H}_{cdr}^{Z} =& \sigma(\bm{u}_{g3_{i}^{*}}^{Z}\bm{W}_{s}^{Z}+ \bm{b}_{s}^{Z}+ \bm{u}_{other_{i}}^{Z}\bm{W}_{o}^{Z}+ \bm{b}_{o}^{Z}),\vspace{1.5ex}\nonumber\\
\bm{u}_{g3_{i}^{'}}^{Z} =& \tanh{((1-\bm{H}_{cdr}^{Z}) \odot \bm{u}_{g3_{i}^{*}}^{Z}+ \bm{H}_{cdr}^{Z} \odot \bm{u}_{other_{i}}^{Z})},
\label{gate_2}
\end{align}
\end{small}
where $\{\bm{W}_{s}^{Z}, \bm{W}_{o}^{Z}\}$ $\in \mathbb{R}^{D_{cgm}\times D_{cgm}}$ and $\{\bm{b}_{s}^{Z}, \bm{b}_{o}^{Z}\}$ $\in \mathbb{R}^{D_{cgm}}$ are the trainable weights and biases. $\sigma(\cdot)$ is the sigmoid function and  $\odot$ is the Hadamard product. The representation of $u_{i}^{Z}$ after inter node matching component is obtained as:
\begin{small}
\begin{equation}
\begin{array}{cc}
\bm{u}_{g3_{i}}^{Z} = \bm{u}_{g3_{i}^{'}}^{Z} + \bm{u}_{g2_{i}}^{Z}.
\label{add_cross}
\end{array}
\end{equation}
\end{small}
\textbf{Similar inter node matching processes are operated on domain $\bar{Z}$.} 
\subsection{Intra Node Complementing Module}
Intra-to-inter node matching module complements the user's latent interests by transferring the information within and cross domains, but the insufficiency of the user representation still remains due to their observed sparse historical behaviors. In order to further tackle this issue,
we propose a node complementing module to correct the biased representations before ranking recommendation tasks. Concretely, we complement the potential missing interactions by measuring the similarity between the user and item representations (i.e., user-item matching procedure) and then generate \textit{virtual} link strength for each domain. As for a user-item pair ($u_{i}^{Z}$,$v_{j}^{Z}$), the \textit{virtual} link strength can be calculated as follows.
\vspace{-15pt}

\begin{small}
\begin{align}
\alpha_{u_{i}^{Z}v_{j}^{Z}} = \dfrac{\exp(\bm{u}_{g3_{i}}^{Z}{\bm{v}_{j}^{Z}}^{T})}{\sum\limits_{v_{j}\in \mathcal{N}_{u_{i}^{Z}}}\exp(\bm{u}_{g3_{i}}^{Z}{\bm{v}_{j}^{Z}}^{T})}.
\label{m4alpha}
\end{align}
\end{small}
With the \textit{virtual} link strength, we update the user representation as:

\begin{small}
\begin{equation}
\begin{array}{cc}
\bm{u}_{g4_{i}}^{Z} = \bm{u}_{g3_{i}}^{Z} + \sum\limits_{v_{j}\in \mathcal{N}_{u_{i}^{Z}}}\alpha_{u_{i}^{Z}v_{j}^{Z}}\bm{v}_{j}^{Z} \bm{W}_{ref}^{Z} + \bm{b}_{ref}^{Z},
\label{m4update}
\end{array}
\end{equation}
\end{small}
where $\bm{W}_{ref}^{Z}$ $\in$ $\mathbb{R}^{D_{cgm} \times D_{ref}}$ and $\bm{b}_{ref}^{Z}$ $\in \mathbb{R}^{D_{ref}}$ are the trainable parameters of the node complementing operation. $D_{ref}$ is the transformation dimension. \textbf{Similar intra node complementing processes are operated on domain $\bar{Z}$.} 

\subsection{Prediction Layer}
After obtaining user/item representations, we construct a prediction layer to estimate the user’s preference towards the target item as:
\begin{small}
\begin{equation}
\begin{array}{cc}
\hat{y}_{u_{i},v_{j}}^{Z}=\sigma( \mathrm{MLPs} (\bm{u}_{g4_{i}}^{Z}||\bm{v}_{j}^{Z}))
\label{predict1}
\end{array}
\end{equation}
\end{small}
where MLPs are the stacked MLP layers with the input of the concatenation of the user and item embeddings. $\sigma$ denotes the sigmoid function. \textbf{The prediction layer in the domain $\bar{Z}$ is similar.} 

\subsection{Companion Objective and Loss Function}
\wujiang{Inspired by \cite{xu2022recursive,lee2015deeply}, we insert companion objectives into each key module to regularize the embedding learning and expedite model convergence during training. Given the user-item pairs corresponding to each key component as mentioned above, i.e., ($\bm{u}_{i}^{Z}$, $\bm{v}_{j}^{Z}$), ($\bm{u}_{g1_{i}}^{Z}$, $\bm{v}_{j}^{Z}$), ($\bm{u}_{g2_{i}}^{Z}$, $\bm{v}_{j}^{Z}$) and ($\bm{u}_{g3_{i}}^{Z}$, $\bm{v}_{j}^{Z}$), each of them is fed into a shared prediction layer and we can get the corresponding prediction outputs as $\hat{y}_{g0}^{Z}$, $\hat{y}_{g1}^{Z}$, $\hat{y}_{g2}^{Z}$ and $\hat{y}_{g3}^{Z}$ according to Eq. \ref{predict1}.
In this work, we adopt the Binary Cross Entropy (BCE) loss for the companion objectives. The common definition of the BCE loss can be formulated as:}
\begin{small}
\begin{equation}
\textcolor{black}{\ell(\hat{y},y) = - [ y\log \hat{y} + (1-y)\log(1-\hat{y})].}
\label{bce}
\end{equation}
\end{small}
\textcolor{black}{$\hat{y}$ is the prediction result and $y$ represents the ground-truth label. The companion objectives can be written as follows:}

\begin{small}
\begin{align}
\textcolor{black}{\mathcal{L}_{CO}^{Z} =} & \textcolor{black}{\sum_{\substack{u_{i}\in \mathcal{U}^Z,v_{j}\in \mathcal{V}^Z}}	
 \Big[w_1\ell(\hat{y}_{g0_{u_{i}v_{j}}}^{Z},y_{u_{i}v_{j}}^{Z}) 
 +w_2\ell(\hat{y}_{g1_{u_{i}v_{j}}}^{Z},y_{u_{i}v_{j}}^{Z})}\nonumber\\
 & \textcolor{black}{+ w_3\ell(\hat{y}_{g2_{u_{i}v_{j}}}^{Z},y_{u_{i}v_{j}}^{Z})+ 
 w_4\ell(\hat{y}_{g3_{u_{i}v_{j}}}^{Z},y_{u_{i}v_{j}}^{Z}) \Big]},
\label{CO_lossX}
\end{align}
\end{small}
\wujiang{where $y_{u_{i}v_{j}}^{Z}$ is the ground-truth label for a real interaction between $u_i$ and $v_j$ in domain $Z$, and $w_{1,2,3,4}$ is the static or dynamically computed weight per term. Besides, except for the above companion objectives loss, the model final prediction loss is written as:}
\vspace{-15pt}

\begin{small}
\begin{align}
\textcolor{black}{\mathcal{L}_{cls}^{Z}= \sum_{\substack{u_{i}\in \mathcal{U}^Z,\\v_{j}\in \mathcal{V}^Z}}	
  \ell(\hat{y}_{u_{i}v_{j}}^{Z},y_{u_{i}v_{j}}^{Z}).}
\label{cls_loss}
\end{align}
\end{small}
\vspace{-15pt}

\wujiang{\textbf{The companion losses and final prediction loss for domain $\bar{Z}$ could be obtained in a similar way.} Finally, the overall loss could be obtained as: }
\begin{small}
\begin{equation}
\textcolor{black}{\mathcal{L}_{total}=w_5\mathcal{L}_{CO}^{Z}+w_6\mathcal{L}_{CO}^{\bar{Z}}+w_7\mathcal{L}_{cls}^{Z}+w_8\mathcal{L}_{cls}^{\bar{Z}}.}
\label{total_loss}
\end{equation}
\end{small}
\wujiang{where $w_{5,6,7,8}$ are tradeoff parameters.}

\subsection{Theoretical Analysis of Model Stability}
To provide a theoretical insight into our model performance, we conduct an essential stability analysis in this section. Following the works \cite{keriven2020convergence,gama2020stability,agarwal2022probing}, the stability of one model could be defined as:
\begin{definition}
\textcolor{black}
{Given user node $u$ and item node $v$ within a graph, a GNN model framework $\Phi$ is said to be stable if:
\begin{equation}
    \|z_{u,v}-z_{u',v}\|_2\leq\gamma\|x_u-x_u'\|,
\end{equation}
where $u'$ denotes the user node $u$ with perturbations.} $z_{u,v}$ represents the predicted possible interactions of $u$ and $v$ by framework $\Phi$. $x_u$ and $x_u'$ are the node embeddings for $u$ and $u'$. $\gamma$ denotes the Lipschitz constant. 
\end{definition}
\textbf{To derive the upper bound of our model instability,} we compress our model into three layers, i.e., a heterogeneous graph encoder layer (first layer), a fully connected homogeneous graph encoder layer (second layer), and a prediction layer (third layer). The representations of $u$ after the first and second layer are formulated as: 

\begin{small}
\begin{align}
    h_u^1=&sp(\bm{W}_a^1 x_u+\frac{1}{n}\bm{W}_n^1\sum_{v\in \mathcal{N}_u}x_v+\bm{b}_1), \vspace{0.8ex}\nonumber\\
    h_u^2=&sp(\bm{W}_a^2 h_u^1+\frac{1}{N-1}\bm{W}_n^2\sum_{v\in \mathcal{G}\setminus u}h_v^1+\bm{b}_2), \vspace{0.8ex}
\end{align}
\end{small}

Similarly, we can get the representation $h_v^2$ for node $v$. Then

\begin{equation}
    z_{u,v}=\text{softmax}(\bm{W}_{a}^{3}(h_u^2\|h_v^2)+\bm{b}_3).
\end{equation}

where $\bm{W}_a^1$, $\bm{W}_n^1$, $\bm{W}_a^2$, $\bm{W}_n^2$ and $\bm{W}_a^3$ are the transformation matrix, $\bm{b}_1$, $\bm{b}_2$, $\bm{b}_3$ are the bias, $\mathcal{N}_u$ denotes the first-order neighbor of the user $u$, $\mathcal{G}\setminus u$ denotes the users in the graph $G$ expect user $u$, $n$ denotes the number of neighborhood of $u$, $N$ is the total number of node in graph. $sp$ denotes the softplus activation function, which is a smooth approximation of the ReLU function. Similar operations are also operated on $u'$. Consequently, we can get:

\begin{small}
\begin{align}
    \|z_{u,v}&-z_{u',v}\|_2=\|\text{softmax}(s_{u,v})-\text{softmax}(s_{u',v})\|_2\nonumber\\
    &\leq\mathcal{C}_{sf}\|\bm{W}_{a}^{3}\|_2\|sp(\bm{W}_a^2 h_u^1+\frac{1}{N-1}\bm{W}_n^2\sum_{v\in \mathcal{G}\setminus u}h_v^1+\bm{b}_2)\nonumber\\
    &-sp(\bm{W}_a^2 h_u'^1+\frac{1}{N-1}\bm{W}_n^2\sum_{v\in \mathcal{G}\setminus u}h_{v'}^1+\bm{b}_2)\|_2\nonumber\\
    &\leq\mathcal{C}_{sf}\mathcal{C}_{sp}\|\bm{W}_{a}^{3}\|_2\|(\|\bm{W}_a^2\|_2\|h_u^1-h_u'^1\|_2\nonumber\\
    &+\frac{1}{N-1}\|\bm{W}_n^2\|_2\|\sum_{v\in \mathcal{G}\setminus u}h_{v}^1-\sum_{v\in \mathcal{G}\setminus u}h_{v'}^1\|_2)\|_2,
\end{align}
\end{small}

where $\mathcal{C}_{sf}$ and $\mathcal{C}_{sp}$ represents the Lipschitz constant for the softmax and softplus function respectively. 
Since 

\begin{small}
\begin{align}
    \|h_u^1-h_u'^1\|_2=&\|sp(\bm{W}_a^1 x_u+\frac{1}{n}\bm{W}_n^1\sum_{v\in \mathcal{N}_u}x_v+\bm{b}_1)\nonumber\\
    &-sp(\bm{W}_a^1 x_u'+\frac{1}{n}\bm{W}_n^1\sum_{v\in \mathcal{N}_u}x_v+\bm{b}_1)\|_2\nonumber\\
    \leq&\mathcal{C}_{sp}\|\bm{W}_a^1\|_2\|x_u-x_u'\|_2,
\end{align}
\end{small}
For $v_i \notin \mathcal{N}_u$, we can get $h_{v_i'}^1-h_{v_i}^1=0$. For $v_j \in \mathcal{N}_u$,

\begin{small}
\begin{align}
    \|h_{v_j'}^1&-h_{v_j}^1\|_2\nonumber\\
    =&\|sp(\bm{W}_a^1 x_{v_j}+\frac{1}{n_j}\bm{W}_n^1\sum_{k\in \mathcal{N}_{v_j}\setminus u}x_k+ \frac{1}{n_j}\bm{W}_n^1 x_u+\bm{b}_1)\nonumber\\
    &-sp(\bm{W}_a^1 x_{v_j}+\frac{1}{n_j}\bm{W}_n^1\sum_{k\in \mathcal{N}_{v_j}\setminus u}x_k+ \frac{1}{n_j}\bm{W}_n^1 x_u'+\bm{b}_1)\|_2\nonumber\\
    \leq&\frac{1}{n_j}\mathcal{C}_{sp}\|\bm{W}_n^1\|_2\| x_u-x_u'\|_2,
\end{align}
\end{small}
where $n_j$ is the number of neighborhood of $v_j$. \textbf{Then, we can get the instability upper bound of our model as:}
\begin{small}
\begin{align}
    \|z_{u,v}- &z_{u',v}\|_2
    \leq\mathcal{C}_{sf}\mathcal{C}_{sp}^2\|\bm{W}_{a}^{3}\|_2(\|\bm{W}_a^2\|_2\|\bm{W}_a^1\|_2\nonumber\\
    &+\frac{\sum_{v_j \in \mathcal{N}_u}\frac{1}{n_j}}{N-1}\|\bm{W}_n^2\|_2\|\bm{W}_n^1\|_2)\|x_u-x_u'\|_2
\label{thom}
\end{align}
\end{small}
\textcolor{black}{Noting that an appropriate instability upper bound is essential for one model's robustness (cannot be too large) and discernibility (cannot be too small). As shown in Eq. \ref{thom}, we observe the model instability upper bound is quite correlated with the norm of transformation matrix. In ideal cases, each user/item should have distinct learnable transformation matrices to get an appropriate instability upper bound for the model. However, too many learnable transformation matrices would result in model parameter explosion and is quite unpractical. Thus, in this paper, we distinguish head and tail users, according to their number of neighborhood and utilize different learnable transformation matrices instead of common one.}

\section{Experiments}
In this section, we first present the experimental settings, including the datasets, evaluation metrics and comparison methods. Then, we conduct several detailed experiments to answer the following questions (RQs):
\begin{itemize}
\item{
\textbf{RQ1}: \textcolor{black}{How does NMCDR perform on (few) partially overlapped multi-target CDR scenarios compared with the state-of-the-art methods?}
}
\item{
\textbf{RQ2}: How do the different modules of NMCDR contribute to the performance gain of our method?}
\item{
\textbf{RQ3}: How do different hyperparameter settings of NMCDR influence the recommendation performance?}
\end{itemize}

\subsection{Experimental Setting}
\subsubsection{Datasets} 
We conduct experiments on four tasks derived from a public and a real-world industrial dataset.
Following existing researches \cite{li2020ddtcdr,ouyang2020minet,li2021dual,zhu2022personalized,cui2020herograph}, we evaluate our method on Amazon\footnote{http://jmcauley.ucsd.edu/data/amazon/index\_2014.html} \cite{he2016ups} datasets, \textcolor{black}{which consist of 24 disjoint item domains and we select 3 pairs of domains to formulate three tasks, i.e., ``Music-Movie'', ``Cloth-Sport'' and ``Phone-Elec''.} Besides, we conduct another task on a large-scale financial CDR dataset, which is collected from traffic logs of the online recommender system of MYbank of Ant Group\footnote{https://www.antgroup.com/en}. The financial dataset describes users' interactions in financial products such as debit, trust, i.e., ``Loan-Fund''. The concrete statistics of each task are summarized in Table \ref{anylis}.


\begin{table}[!h]
\footnotesize
\setlength{\abovecaptionskip}{0pt}
\setlength{\belowcaptionskip}{5pt}
\captionsetup{font={footnotesize}}
\caption{Statistics on the Amazon and MYbank datasets.}
\label{anylis}
\begin{threeparttable} 
\setlength\tabcolsep{3pt}{
{
\begin{tabular}{cc|cc|cc|c}
\toprule
\multicolumn{2}{c|}{\textbf{Dataset}}     & \textbf{Users}       & \textbf{Items}       & \textbf{Ratings}   & \textbf{\#Overlapping}           &  \textbf{Density} \\ \midrule 
\multirow{2}{*}{Amazon} & Music & 50,841  & 43,858 & 713,740   & \multirow{2}{*}{15,081} & 0.03\% \\
                        & Movie & 87,875  & 38,643 & 1,184,889 &                         & 0.03\%  \\ \midrule 
\multirow{2}{*}{Amazon} & Cloth & 27,519  & 9,481  & 161,010   & \multirow{2}{*}{16,337} & 0.06\%  \\
                        & Sport & 107,984 & 40,460 & 851,553   &                         & 0.02\%  \\ \midrule 
\multirow{2}{*}{Amazon} & Phone & 41,829  & 17,943 & 194,121   & \multirow{2}{*}{7,857}  & 0.03\%  \\
                        & Elec  & 27,328  & 12,655 & 170,426   &                         & 0.05\% \\ \midrule 
\multirow{2}{*}{MYbank} & Loan  & 147,837 & 1,488  & 304,409   & \multirow{2}{*}{6,530}  & 0.14\%  \\
                        & Fund  & 65,257  & 1,319  & 86,281    &                         & 0.10\%  \\
\bottomrule
\end{tabular}
}}
\begin{tablenotes}    
        \footnotesize            
        \item \#Overlapping denotes the number of overlapping users across domains. 
      \end{tablenotes}
\end{threeparttable}
\end{table}
\vspace{-10pt}
\subsubsection{Evaluation Metrics}
\textcolor{black}{To verify NMCDR's capability of handling partially overlapping multi-target CDR tasks, we vary the overlapping ratio $\mathcal{K}_{u}$ of each dataset in $\{0.1\%, 1\%, 10\%, 50\%, 90\%\}$. Different overlapping ratios indicate that different numbers of common users are shared across domains. For example, in Amazon "Music-Movie" dataset with $\mathcal{K}_{u}=10\%$, the number of the overlapped users is calculated like $15,081*0.1 = 1508$.}
Following common practice in previous CDR literature \cite{he2017neural,rendle2012bpr,cao2022cross} , \textcolor{black}{we utilize the \textit{leave-one-out} technique to evaluate the performance of the developed model}. Meanwhile, we follow the above works and randomly sample 199 negative items (i.e., items are not interacted by the user) along with 1 positive item (i.e., ground-truth interaction) to form the recommendation candidates to conduct the ranking test. Based on the ranking results, we utilize the typical $top$-$N$ metrics normalized discounted cumulative gain (NDCG@10), and hit rate (HR@10) to evaluate the model performance, which are frequently used in the CDR scenarios \cite{zhu2021unified, li2020ddtcdr, zhu2019dtcdr}. For all the metrics, higher values indicate better performance.


\subsubsection{Comparison Methods}
\textcolor{black}{We quantitatively compare NMCDR against several state-of-the-art methods which can be divided into three classes.} 

\begin{table*}[h!]
\scriptsize
\captionsetup{font={footnotesize}}
\caption{Experimental results (\%) on the bi-directional Music-Movie CDR scenario with different user overlapped ratio.}
\label{musicmoviecompare}
\setlength\tabcolsep{1.6pt}{
{
\begin{tabular}{cccccccccccccccccccccccccc}
\toprule
\multirow{4}{*}{\bf Methods} & \multicolumn{10}{c}{\textbf{Music-domain recommendation }} & \multicolumn{10}{c}{\textbf{Movie-domain recommendation }} \\
\cmidrule(r){2-21}
& \multicolumn{2}{c}{\ $\mathcal{K}_{u}$=0.1\%} & \multicolumn{2}{c}{\ $\mathcal{K}_{u}$=1\%} 
& \multicolumn{2}{c}{\ $\mathcal{K}_{u}$=10\%} & \multicolumn{2}{c}{\ $\mathcal{K}_{u}$=50\%} & \multicolumn{2}{c}{\ $\mathcal{K}_{u}$=90\%} 
& \multicolumn{2}{c}{\ $\mathcal{K}_{u}$=0.1\%} & \multicolumn{2}{c}{\ $\mathcal{K}_{u}$=1\%}  & \multicolumn{2}{c}{\ $\mathcal{K}_{u}$=10\%} & \multicolumn{2}{c}{\ $\mathcal{K}_{u}$=50\%} & \multicolumn{2}{c}{\ $\mathcal{K}_{u}$=90\%}\\ \cmidrule(r){2-3}\cmidrule(r){4-5}\cmidrule(r){6-7}\cmidrule(r){8-9}\cmidrule(r){10-11}\cmidrule(r){12-13}
\cmidrule(r){14-15}\cmidrule(r){16-17}\cmidrule(r){18-19}\cmidrule(r){20-21}
&NDCG    &HR     
&NDCG    &HR   &NDCG    &HR    &NDCG    &HR 
&NDCG    &HR   &NDCG    &HR    &NDCG    &HR   
&NDCG    &HR   &NDCG    &HR    &NDCG    &HR \\
\midrule

LR \cite{richardson2007predicting} 
& 5.25
& 9.31
& 5.78
& 10.03
&  5.92
&  11.40
&  7.36
&  14.41
&  9.74
&  18.58
& 31.36
& 47.08
& 31.41
& 47.01
&  31.61
&  47.62
&  31.66
&  47.76
&  31.64
&  47.66 \\

BPR \cite{rendle2012bpr} 
& 2.97
& 6.63
& 2.92
& 6.77
&  2.67
&  5.76
&  2.79
&  6.15
&  2.92 
&  6.26 
& 21.63
& 35.59
& 21.65
& 35.61
&  21.79 
&  35.78 
&  22.00 
&  36.09 
&  21.97 
&  36.14 \\

NeuMF \cite{he2017neural} 
& 4.86
& 9.17
& 5.01
& 9.78
&  5.07 
&  9.87 
&  5.58 
&  11.18 
&  6.00 
&  11.93
& 28.79
& 43.27
& 28.96
& 42.84
&  29.02 
&  43.58 
&  29.32 
&  44.16 
&  29.21 
&  43.91    \\

\midrule

MMoE \cite{ma2018modeling}  
& 6.60
& 12.83
& 6.85
& 14.25
&  6.95 
&  14.69 
&  9.02 
&  18.30 
&  10.44 
&  20.70 
& 30.20
& 48.54
& 31.15
& 47.12
&  31.31 
&  47.77 
&  31.32 
&  47.84 
&  31.80 
&  48.07  \\

PLE \cite{tang2020progressive} 
& 6.66
& 13.12
& 6.89
& 14.26
&  7.25 
&  14.60 
&  9.00 
&  17.64 
&  10.08 
&  19.78 
& 31.72
& 47.47
& 31.83
& 47.46
&  31.96 
&  47.89 
&  32.04 
&  47.89 
&  32.02 
&  48.03  \\

\midrule
CoNet \cite{hu2018conet}  
& 7.03
& 14.10
& 7.26
& 14.40
&  7.48 
&  15.24 
&  9.61 
&  19.47 
&  10.19 
&  20.75 
& 31.06
& 47.07
& 31.26
& 47.24
&  31.30 
&  47.42 
&  31.40 
&  47.55 
&  31.37 
&  47.51  \\

MiNet \cite{ouyang2020minet} 
& 5.19
& 11.42
& 5.67
& 11.85
&  6.24 
&  12.43 
&  8.84 
&  17.16 
&  9.37 
&  17.69 
& 29.95
& 44.78
& 30.22
& 45.25
&  29.85 
&  45.01 
&  29.58 
&  44.84 
&  29.67 
&  45.13  \\

GA-DTCDR \cite{zhu2020graphical}  
& 7.03
& 14.03
& 7.17
& 14.53
&  7.26 
&  14.60 
&  9.54 
&  19.17 
&  10.16 
&  19.97 
& 31.56
& 47.36
& 31.61
& 47.41
&  31.70 
&  47.63 
&  31.90 
&  47.77 
&  31.85 
&  47.81 \\

\midrule

DML \cite{li2021dual}  
& 6.81
& 13.08
& 7.32
& 13.54
&  7.99 
&  15.58 
&  9.58 
&  18.66 
&  10.55 
&  20.33 
& 26.36
& 40.84
& 27.06
& 41.47
&  27.44 
&  41.63 
&  27.36 
&  41.76 
&  27.42 
&  41.86 \\

HeroGraph \cite{cui2020herograph}  
& 6.59
& 13.40
& 7.44
& 13.89
&  7.02 
&  14.49 
&  9.15 
&  18.55 
&  10.34 
&  20.33 
& \underline{32.05}
& \underline{48.14}
& \underline{32.22}
& \underline{48.38}
&  \underline{32.16}
&  \underline{48.40}
&  \underline{32.23}
&  \underline{48.52}
&  \underline{32.18}
&  \underline{48.43}\\

PTUPCDR \cite{zhu2022personalized}
& \underline{7.60}
& \underline{14.95}
& \underline{7.75}
& \underline{15.23}
&  \underline{8.28}
&  \underline{16.58}
&  \underline{9.89}
&  \underline{20.08}
&  \underline{10.97}
&  \underline{21.31}
& 31.80
& 47.31
& 31.92
& 47.65
&  31.92 
&  47.84 
&  31.90 
&  47.94 
&  31.93 
&  47.96 \\

\midrule

NMCDR  
& \textbf{8.29}
& \textbf{16.28}
& \textbf{8.43}
& \textbf{16.52}
& \textbf{ 8.50 }
& \textbf{ 17.00 }
& \textbf{ 11.26 }
& \textbf{ 21.58 }
& \textbf{ 12.28 }
& \textbf{ 23.19 }
& \textbf{33.39}
& \textbf{50.22}
& \textbf{33.57}
& \textbf{50.67}
& \textbf{ 33.70 }
& \textbf{ 50.91 }
& \textbf{ 33.96 }
& \textbf{ 51.13 }
& \textbf{ 33.94 }
& \textbf{ 51.12 } \\

Improvement(\%) & 9.08 & 8.90 & 8.77 & 8.47 &2.66 & 2.53 & 13.85 & 7.47 & 11.94  & 8.82
&  4.18 & 4.32 & 4.19 &  4.73
& 4.79  & 5.19 & 5.37 & 5.38 & 5.47 & 5.55 \\
\bottomrule
\end{tabular}
}}
\end{table*}

\begin{table*}[h!]
\scriptsize
\captionsetup{font={footnotesize}}
\caption{Experimental results (\%) on the bi-directional Cloth-Sport CDR scenario with different user overlapped ratio.}
\label{clothsportcompare}
\setlength\tabcolsep{1.2pt}{
{
\begin{tabular}{cccccccccccccccccccccccccc}
\toprule
\multirow{4}{*}{\bf Methods} & \multicolumn{10}{c}{\textbf{Cloth-domain recommendation }} & \multicolumn{10}{c}{\textbf{Sport-domain recommendation }} \\
\cmidrule(r){2-21}
& \multicolumn{2}{c}{\ $\mathcal{K}_{u}$=0.1\%} & \multicolumn{2}{c}{\ $\mathcal{K}_{u}$=1\%} 
& \multicolumn{2}{c}{\ $\mathcal{K}_{u}$=10\%} & \multicolumn{2}{c}{\ $\mathcal{K}_{u}$=50\%} & \multicolumn{2}{c}{\ $\mathcal{K}_{u}$=90\%} 
& \multicolumn{2}{c}{\ $\mathcal{K}_{u}$=0.1\%} & \multicolumn{2}{c}{\ $\mathcal{K}_{u}$=1\%}  & \multicolumn{2}{c}{\ $\mathcal{K}_{u}$=10\%} & \multicolumn{2}{c}{\ $\mathcal{K}_{u}$=50\%} & \multicolumn{2}{c}{\ $\mathcal{K}_{u}$=90\%}\\ \cmidrule(r){2-3}\cmidrule(r){4-5}\cmidrule(r){6-7}\cmidrule(r){8-9}\cmidrule(r){10-11}\cmidrule(r){12-13}
\cmidrule(r){14-15}\cmidrule(r){16-17}\cmidrule(r){18-19}\cmidrule(r){20-21}
&NDCG    &HR     
&NDCG    &HR   &NDCG    &HR    &NDCG    &HR 
&NDCG    &HR   &NDCG    &HR    &NDCG    &HR   
&NDCG    &HR   &NDCG    &HR    &NDCG    &HR \\
\midrule

LR \cite{richardson2007predicting}  
& 5.02
& 11.03
& 5.64
& 11.58
&  6.32 
&  12.40 
&  6.65 
&  13.13 
&  7.16 
&  14.18
& 9.24 
& 18.39
& 10.01
& 19.14
&  10.79 
&  20.12 
&  11.28 
&  21.15 
&  11.45 
&  21.36  \\

BPR \cite{rendle2012bpr} 
& 2.52
& 5.65
& 2.60
& 5.70
&  2.70 
&  5.87 
&  2.66 
&  5.93 
&  2.74 
&  5.93 
& 2.38
& 5.13
& 2.44
& 5.33
&  2.64 
&  5.88 
&  2.74 
&  6.04 
&  2.79 
&  6.04 \\

NeuMF \cite{he2017neural} 
& 2.88
& 7.02
& 3.48
& 7.65
&  4.26 
&  8.75 
&  4.35 
&  8.82 
&  4.35 
&  9.16 
& 6.19
& 11.45
& 6.43
& 12.43
&  6.71 
&  12.96 
&  7.09 
&  13.62 
&  7.52 
&  14.41 \\

\midrule
MMoE \cite{ma2018modeling}  
& 6.03
& 12.30
& 6.10
& 12.46
&  6.20 
&  12.87 
&  6.65 
&  13.73 
&  7.03 
&  14.50 
& 9.89
& 18.99
& 9.97
& 19.08
&  10.43 
&  19.84 
&  10.89 
&  20.81 
&  11.39 
&  21.76 \\

PLE \cite{tang2020progressive} 
&   5.85  
&   11.62  
&   6.02  
&   11.85  
&  6.29 
&  12.51 
&  7.00 
&  14.01 
&  7.15 
&  14.35 
&   9.98  
&   18.35  
&   10.01  
&   18.44  
&  10.49 
&  19.68 
&  11.31 
&  20.87 
&  11.39 
&  21.05 \\

\midrule
CoNet \cite{hu2018conet} 
&   6.02  
&   12.06  
&   6.13  
&   12.52  
&  6.26 
&  12.85 
&  6.88 
&  14.02 
&  7.33 
&  14.79
&   9.59  
&   18.30  
&   9.68  
&   18.49  
&  9.84 
&  18.63 
&  10.84 
&  20.52 
&  11.23 
&  21.35 \\

MiNet \cite{ouyang2020minet} 
&   5.07  
&   10.40  
&   5.24  
&   10.61  
&  5.41 
&  10.87 
&  6.17 
&  12.51 
&  6.66 
&  13.35 
&   8.37  
&   16.05  
&   8.62  
&   16.62  
&  8.84 
&  16.98 
&  9.72 
&  18.30 
&  10.58 
&  19.96  \\

GA-DTCDR \cite{zhu2020graphical} 
&   5.61  
&   12.13  
&   5.68  
&   12.28  
&  6.22 
&  12.90 
&  7.04 
&  14.06 
&  \underline{7.59}
&  14.85 
&   \underline{10.71}
&   \underline{20.28}
&   10.75  
&   \underline{20.34}
&  10.91 
&  20.55 
&  11.63 
&  21.86 
&  \underline{12.25}
&  \underline{22.96}\\

\midrule

DML \cite{li2021dual}  
&   5.37  
&   10.63  
&   5.44  
&   10.90  
&  5.59 
&  11.10 
&  6.31 
&  12.57 
&  6.55 
&  12.96 
&   6.51  
&   12.42  
&   6.53  
&   12.49  
&  6.62 
&  12.73 
&  7.05 
&  13.47 
&  7.75 
&  14.99 \\

HeroGraph \cite{cui2020herograph}  
&   6.21  
&   12.30  
&   6.34  
&   12.53  
&  6.37 
&  12.75 
&  7.06 
&  13.90 
&  7.51 
&  14.75 
&   10.45  
&   19.53  
&   10.52  
&   19.91  
&  11.06 
&  20.74 
&  11.77 
&  21.73 
&  12.24 
&  22.75 \\

PTUPCDR \cite{zhu2022personalized} 
&   \underline{6.22}
&   \underline{13.07}
&   \underline{6.63}
&   \underline{13.24}
&  \underline{6.79}
&  \underline{13.76}
&  \underline{7.36}
&  \underline{14.78}
&  7.58 
&  \underline{15.52}
&   10.66  
&   19.88  
&   \underline{10.91}
&   20.33  
&  \underline{11.14}
&  \underline{20.77}
&  \underline{11.79}
&  \underline{22.20}
&  12.18 
&  22.95 \\

\midrule

NMCDR 
& \textbf{  8.40  }
& \textbf{  16.57  }
& \textbf{  8.50  }
& \textbf{  16.63  }
& \textbf{ 8.87 }
& \textbf{ 17.73 }
& \textbf{ 9.26 }
& \textbf{ 18.33 }
& \textbf{ 9.54 }
& \textbf{ 19.05 }
& \textbf{  13.52  }
    & \textbf{  25.36  }
    & \textbf{  13.79  }
    & \textbf{  25.53  }
& \textbf{ 14.06 }
& \textbf{ 26.15 }
& \textbf{ 14.91 }
& \textbf{ 27.54 }
& \textbf{ 15.17 }
& \textbf{ 28.10 } \\

Improvement(\%)  &35.05 & 26.78 & 28.21    &25.60      &30.63    & 28.85    & 25.82    & 24.02    & 25.69    & 22.74    & 26.24     & 25.05    & 26.40    &25.52   
& 26.21    & 25.90    & 26.46    & 24.05    & 23.84    & 22.39    \\
\bottomrule
\end{tabular}
}}
\end{table*}

\begin{table*}[h!]
\scriptsize
\captionsetup{font={footnotesize}}
\caption{Experimental results (\%) on the bi-directional Phone-Elec CDR scenario with different user overlapped ratio.}
\label{phoneeleccompare}
\setlength\tabcolsep{1.8pt}{
{
\begin{tabular}{cccccccccccccccccccccccccc}
\toprule
\multirow{4}{*}{\bf Methods} & \multicolumn{10}{c}{\textbf{Phone-domain recommendation }} & \multicolumn{10}{c}{\textbf{Elec-domain recommendation }} \\
\cmidrule(r){2-21}
& \multicolumn{2}{c}{\ $\mathcal{K}_{u}$=0.1\%} & \multicolumn{2}{c}{\ $\mathcal{K}_{u}$=1\%} 
& \multicolumn{2}{c}{\ $\mathcal{K}_{u}$=10\%} & \multicolumn{2}{c}{\ $\mathcal{K}_{u}$=50\%} & \multicolumn{2}{c}{\ $\mathcal{K}_{u}$=90\%} 
& \multicolumn{2}{c}{\ $\mathcal{K}_{u}$=0.1\%} & \multicolumn{2}{c}{\ $\mathcal{K}_{u}$=1\%}  & \multicolumn{2}{c}{\ $\mathcal{K}_{u}$=10\%} & \multicolumn{2}{c}{\ $\mathcal{K}_{u}$=50\%} & \multicolumn{2}{c}{\ $\mathcal{K}_{u}$=90\%}\\ \cmidrule(r){2-3}\cmidrule(r){4-5}\cmidrule(r){6-7}\cmidrule(r){8-9}\cmidrule(r){10-11}\cmidrule(r){12-13}
\cmidrule(r){14-15}\cmidrule(r){16-17}\cmidrule(r){18-19}\cmidrule(r){20-21}
&NDCG    &HR     
&NDCG    &HR   &NDCG    &HR    &NDCG    &HR 
&NDCG    &HR   &NDCG    &HR    &NDCG    &HR   
&NDCG    &HR   &NDCG    &HR    &NDCG    &HR \\
\midrule

LR \cite{richardson2007predicting}  
& 4.12
& 7.83
& 4.54
& 8.75
&  5.96 
&  12.03 
&  13.06 
&  23.29 
&  15.03 
&  26.58 
& 19.67
& 31.43
& 19.99
& 31.91
&  19.98 
&  32.48 
&  20.88 
&  33.83 
&  21.29 
&  34.47 \\

BPR \cite{rendle2012bpr} 
& 2.49
& 5.22
& 2.56
& 5.32
&  2.55 
&  5.58 
&  2.67 
&  5.84 
&  3.10 
&  6.72
& 8.39
& 15.35
& 8.47
& 15.47
&  8.66 
&  15.76 
&  9.80 
&  17.47 
&  10.79 
&  19.07 \\

NeuMF \cite{he2017neural} 
& 3.45
& 6.73
& 3.54
& 7.07
&  4.01 
&  8.34 
&  7.79 
&  14.36 
&  10.40 
&  18.65 
& 15.82
& 25.25
& 16.04
& 26.12
&  16.27 
&  26.17 
&  17.12 
&  27.43 
&  17.77 
&  28.60 \\

\midrule
MMoE \cite{ma2018modeling}  
& 3.95
& 8.71
& 4.18
& 9.05
&  7.54 
&  15.56 
&  13.66 
&  24.85 
&  16.08 
&  28.67 
& 20.16
& 32.07
& 20.27
& 32.83
&  20.85 
&  33.24 
&  21.05 
&  34.05 
&  21.64 
&  34.88 \\

PLE \cite{tang2020progressive}
& 4.24
& 9.13
& 4.82
& 9.92
&  7.27 
&  14.55 
&  13.84 
&  24.94 
&  16.22 
&  28.27 
& 19.95
& 32.61
& 20.32
& 32.73
&  20.75 
&  33.08 
&  21.60 
&  34.44 
&  22.21 
&  35.60 \\

\midrule

CoNet \cite{hu2018conet}  
& 3.93
& 8.16
& 4.02
& 8.46
&  6.88 
&  14.23 
&  13.21 
&  24.26 
&  15.67 
&  28.23 
& 19.65
& 31.57
& 19.77
& 32.13
&  20.20 
&  32.89 
&  21.00 
&  34.10 
&  21.56 
&  35.02 \\

MiNet \cite{ouyang2020minet} 
& 3.56
& 7.58
& 3.66
& 7.70
&  7.22 
&  14.20 
&  13.23 
&  23.51 
&  15.83 
&  27.63 
& 18.22
& 28.61
& 18.99
& 28.64
&  19.30 
&  31.24 
&  19.89 
&  31.90 
&  20.64 
&  33.14 
\\

GA-DTCDR \cite{zhu2020graphical} 
& 3.70
& 7.70
& 4.41
& 9.18
&  7.54 
&  15.14 
&  14.13 
&  25.42 
&  16.36 
&  28.80 
& 20.39
& \underline{32.85}
& 20.55
& 32.90
&  20.75 
&  33.77 
&  21.08 
&  34.08 
&  22.20 
&  35.75  \\

\midrule

DML \cite{li2021dual} 
& \underline{4.56}
& \underline{9.39}
& 4.62
& \underline{9.88}
&  7.08 
&  13.79 
&  12.76 
&  23.21 
&  14.64 
&  26.24 
& 15.70
& 25.59
& 15.72
& 25.66
&  16.09 
&  25.98 
&  16.93 
&  27.38 
&  17.54 
&  28.48 \\

HeroGraph \cite{cui2020herograph}  
& 4.21
& 9.03
& 4.32
& 9.76
& 7.77 
& 15.71 
& 14.22 
& \underline{25.82}
& 16.33 
& 29.20 
& 19.09
& 31.27
& 19.99
& 31.91
& \underline{21.11}
& \underline{34.31}
& 21.19 
& 34.31 
& 21.58 
& 34.84  \\

PTUPCDR \cite{zhu2022personalized} 
& 4.29
& 8.88
& \underline{4.65}
& 9.18
& \underline{8.24}
& \underline{16.30}
& \underline{14.51}
& \underline{25.82}
& \underline{16.84}
& \underline{29.39}
& \underline{20.51}
& 32.73
& \underline{20.60}
& \underline{32.94}
&  20.93 
&  33.89 
& \underline{21.80}
& \underline{35.17}
& \underline{22.31}
& \underline{35.86}\\

\midrule

NMCDR  
& \textbf{6.29}
& \textbf{12.27}
& \textbf{6.46}
& \textbf{12.98}
& \textbf{ 10.82 }
& \textbf{ 20.98 }
& \textbf{ 17.44 }
& \textbf{ 30.87 }
& \textbf{ 19.18 }
& \textbf{ 33.03 }
& \textbf{23.49}
& \textbf{37.61}
& \textbf{23.91}
& \textbf{37.84}
& \textbf{ 24.17 }
& \textbf{ 39.03 }
& \textbf{ 24.45 }
& \textbf{ 39.49 }
& \textbf{ 24.60 }
& \textbf{ 39.84 }\\

Improvement(\%) & 37.93  & 30.67  & 38.92  & 31.38  &31.31 & 28.71 &20.19  & 19.56 & 13.90
& 12.39  
& 14.53  & 14.49  & 16.06  & 14.88
& 14.50  &13.76 & 12.16 & 12.28 & 10.26 & 11.10  \\
\bottomrule
\end{tabular}
}}
\end{table*}

\begin{table*}[h!]
\scriptsize
\captionsetup{font={footnotesize}}
\caption{Experimental results (\%) on the bi-directional Loan-Fund CDR scenario with different user overlapped ratio.}
\label{loanfundcompare}
\setlength\tabcolsep{1.2pt}{
{
\begin{tabular}{cccccccccccccccccccccccccc}
\toprule
\multirow{4}{*}{\bf Methods} & \multicolumn{10}{c}{\textbf{Loan-domain recommendation }} & \multicolumn{10}{c}{\textbf{Fund-domain recommendation }} \\
\cmidrule(r){2-21}
& \multicolumn{2}{c}{\ $\mathcal{K}_{u}$=0.1\%} & \multicolumn{2}{c}{\ $\mathcal{K}_{u}$=1\%} 
& \multicolumn{2}{c}{\ $\mathcal{K}_{u}$=10\%} & \multicolumn{2}{c}{\ $\mathcal{K}_{u}$=50\%} & \multicolumn{2}{c}{\ $\mathcal{K}_{u}$=90\%} 
& \multicolumn{2}{c}{\ $\mathcal{K}_{u}$=0.1\%} & \multicolumn{2}{c}{\ $\mathcal{K}_{u}$=1\%}  & \multicolumn{2}{c}{\ $\mathcal{K}_{u}$=10\%} & \multicolumn{2}{c}{\ $\mathcal{K}_{u}$=50\%} & \multicolumn{2}{c}{\ $\mathcal{K}_{u}$=90\%}\\ \cmidrule(r){2-3}\cmidrule(r){4-5}\cmidrule(r){6-7}\cmidrule(r){8-9}\cmidrule(r){10-11}\cmidrule(r){12-13}
\cmidrule(r){14-15}\cmidrule(r){16-17}\cmidrule(r){18-19}\cmidrule(r){20-21}
&NDCG    &HR     
&NDCG    &HR   &NDCG    &HR    &NDCG    &HR 
&NDCG    &HR   &NDCG    &HR    &NDCG    &HR   
&NDCG    &HR   &NDCG    &HR    &NDCG    &HR \\
\midrule

LR \cite{richardson2007predicting}  
&   47.34  
&   67.59  
&   47.42  
&   67.73  
&  47.65 
&  67.88 
&  47.75 
&  67.82 
&  47.87 
&  68.08 
&   21.97  
&   34.57  
&   22.08  
&   35.65  
&  25.24 
&  36.83 
&  29.70 
&  46.14 
&  31.48 
&  50.98  \\

BPR \cite{rendle2012bpr} 
 &   42.93  
&   62.22  
&   43.07  
&   62.67  
&  43.20 
&  62.99 
&  43.24 
&  63.45 
&  43.37 
&  63.45 
&   3.01  
&   6.28  
&   3.06  
&   6.64  
&  3.21 
&  6.85 
&  2.74 
&  6.51 
&  4.84 
&  10.44  \\

NeuMF \cite{he2017neural} 
 &   46.20  
&   66.66  
&   47.27  
&   67.21   
&  47.74 
&  67.92 
&  48.01 
&  68.19 
&  47.95 
&  68.27 
&   21.53  
&   33.86  
&   21.87  
&   34.07  
&  25.34 
&  37.66 
&  30.78 
&  48.81 
&  30.14 
&  48.73 \\

\midrule
MMoE \cite{ma2018modeling} 
 &   45.23  
&   66.45  
&   45.86  
&   66.88   
&  46.87 
&  67.58 
&  47.81 
&  68.60 
&  47.92 
&  68.55 
&   20.49  
&   34.88  
&   20.59  
&   35.04  
&  20.70 
&  36.53 
&  31.92 
&  52.32 
&  35.84 
&  57.20  \\

PLE \cite{tang2020progressive} 
&   \underline{48.93}
&   \underline{69.01}
&   49.03  
&   \underline{69.28}
&  49.36 
&  69.40 
&  49.31 
&  69.59 
&  49.39 
&  69.79 
&   21.82  
&   \underline{36.09}
&   22.13  
&   36.16  
&  22.91 
&  36.70 
&  33.02 
&  51.40 
&  35.02 
&  55.37 \\

\midrule

CoNet \cite{hu2018conet}  
&   47.85  
&   68.05  
&   48.06  
&   68.25  
&  48.23 
&  68.63 
&  48.37 
&  68.39 
&  48.43 
&  68.65 
&   18.07  
&   29.47  
&   18.60  
&   30.65  
&  20.29 
&  33.03 
&  29.14 
&  49.06 
&  33.97 
&  54.95  \\

MiNet \cite{ouyang2020minet} 
&   47.61  
&   67.59  
&   48.24  
&   68.46  
&  48.84 
&  68.78 
&  48.90 
&  69.01 
&  48.86 
&  69.07 
&   19.89  
&   34.04  
&   21.34  
&   35.82  
&  23.78 
&  37.75 
&  32.18 
&  52.61 
&  34.89 
&  55.91 \\

GA-DTCDR \cite{zhu2020graphical}  
&   45.94  
&   66.51  
&   47.65  
&   68.09  
&  49.20 
&  69.26 
&  49.59 
&  69.86 
&  49.63 
&  \underline{69.94}
&   21.72  
&   32.51  
&   \underline{23.05}
&   34.41  
&  25.40 
&  38.00 
&  33.19 
&  \underline{53.32}
&  \underline{36.60}
&  \underline{57.29}\\

\midrule

DML \cite{li2021dual}  
&   47.12  
&   67.84  
&   47.95  
&   68.56   
&  49.01 
&  \underline{69.77}
&  48.87 
&  69.50 
&  48.84 
&  69.56 
&   21.01  
&   35.75  
&   22.80  
&   \underline{37.35}
&  \underline{25.84}
&  \underline{39.04}
&  32.81 
&  51.44 
&  34.61 
&  54.74 \\

HeroGraph \cite{cui2020herograph}  
&   48.89  
&   68.37  
&   \underline{49.16}
&   68.69  
&  \underline{49.45}
&  69.17 
&  \underline{49.71}
&  69.64 
&  \underline{49.85}
&  69.66 
&   19.07  
&   30.77  
&   19.63  
&   31.44  
&  21.74 
&  33.78 
&  32.23 
&  51.11 
&  35.40 
&  56.41 \\

PTUPCDR \cite{zhu2022personalized} 
&   48.01  
&   68.48  
&   48.32  
&   68.84  
&  49.14 
&  69.32 
&  49.55 
&  \underline{69.91}
&  49.54 
&  69.93
&   \underline{22.13}
&   36.05  
&   22.84  
&   36.83  
&  24.14 
&  37.75 
&  \underline{33.24}
&  53.03 
&  35.61 
&  56.24 \\

\midrule

NMCDR  
& \textbf{  49.47  }
& \textbf{  69.54  }
& \textbf{  49.69  }
& \textbf{  69.84  }
& \textbf{ 49.84 }
& \textbf{ 69.97 }
& \textbf{ 49.89 }
& \textbf{ 69.98 }
& \textbf{ 49.91 }
& \textbf{ 70.06 }
& \textbf{  25.32  }
    & \textbf{  39.47  }
    & \textbf{  25.69  }
    & \textbf{  39.75  }
& \textbf{ 26.38 }
& \textbf{ 40.46 }
& \textbf{ 35.24 }
& \textbf{ 55.03 }
& \textbf{ 37.29 }
& \textbf{ 58.54 }\\

Improvement(\%) & 1.10    & 0.77    & 1.07    & 0.80     & 0.79    & 0.29    & 0.36    & 0.10    & 0.12    & 0.17    
&14.41    & 9.37    & 11.45    & 6.43   
&2.09    & 3.64    & 6.02    & 3.21    & 1.89    & 2.18   \\
\bottomrule
\end{tabular}
}}
\end{table*}

\begin{table*}[h!]
    \scriptsize
    \captionsetup{font={footnotesize}}
    \caption{Experimental results (\%) on the bi-directional Cloth-Sport and Loan-Fund CDR scenarios under different density settings $D_s$.}
    \label{cs_lf_DS}
    \setlength\tabcolsep{0.001pt}{
    {
    \begin{tabular}{cccccccccccccccccccccccccc}
    \toprule
    \multirow{4}{*}{\bf Methods} & \multicolumn{6}{c}{\textbf{Cloth-domain recommendation }} & \multicolumn{6}{c}{\textbf{Sport-domain recommendation }}
    & \multicolumn{6}{c}{\textbf{Loan-domain recommendation }} & \multicolumn{6}{c}{\textbf{Fund-domain recommendation }}\\
    \cmidrule(r){2-25}
    & \multicolumn{2}{c}{$D_s$=10\%} & \multicolumn{2}{c}{$D_s$=50\%} & \multicolumn{2}{c}{$D_s$=70\%} & \multicolumn{2}{c}{$D_s$=10\%} & \multicolumn{2}{c}{$D_s$=50\%} & \multicolumn{2}{c}{$D_s$=70\%}
    & \multicolumn{2}{c}{$D_s$=10\%} & \multicolumn{2}{c}{$D_s$=50\%} & \multicolumn{2}{c}{$D_s$=70\%} & \multicolumn{2}{c}{$D_s$=10\%} & \multicolumn{2}{c}{$D_s$=50\%} & \multicolumn{2}{c}{$D_s$=70\%}
    \\ \cmidrule(r){2-3}\cmidrule(r){4-5}\cmidrule(r){6-7}\cmidrule(r){8-9}\cmidrule(r){10-11}\cmidrule(r){12-13}
    \cmidrule(r){14-15}\cmidrule(r){16-17}\cmidrule(r){18-19}\cmidrule(r){20-21}\cmidrule(r){22-23}\cmidrule(r){24-25}
    &NDCG  &HR &NDCG  &HR  &NDCG  &HR
    &NDCG  &HR &NDCG  &HR  &NDCG  &HR 
    &NDCG  &HR &NDCG  &HR  &NDCG  &HR
    &NDCG  &HR &NDCG  &HR  &NDCG  &HR \\
    \midrule
    
    LR \cite{richardson2007predicting}  &   2.41  
    &   5.38  
    &   2.87  
    &   6.20  
    &   3.21  
    &   6.95  
    &   2.47  
    &   5.42  
    &   2.61  
    &   5.79  
    &   4.20  
    &   8.61
    &  23.30 
    &  32.84 
    &  31.03 
    &  42.66 
    &  38.90 
    &  53.03 
    &  14.54 
    &  23.03 
    &  18.33 
    &  28.43 
    &  19.89 
    &  30.67\\
    
    BPR \cite{rendle2012bpr} &   2.52  
    &   5.61  
    &   2.48  
    &   5.41  
    &   2.45  
    &   5.49  
    &   2.53  
    &   5.63  
    &   2.45  
    &   5.58  
    &   2.66  
    &   5.84 
    &  20.68 
    &  30.70 
    &  27.95 
    &  41.28 
    &  34.04 
    &  50.50 
    &  1.31 
    &  3.11 
    &  1.93 
    &  4.45 
    &  2.50 
    &  5.51 \\
    
    NeuMF \cite{he2017neural} &   2.61  
    &   5.78  
    &   2.74  
    &   5.96  
    &   2.75  
    &   5.96  
    &   2.48  
    &   5.39  
    &   2.68  
    &   5.90  
    &   3.37  
    &   7.02   
    &  23.62 
    &  33.13 
    &  31.41 
    &  45.04 
    &  37.55 
    &  54.66 
    &  15.31 
    &  23.43 
    &  17.47 
    &  25.79 
    &  19.17 
    &  27.84\\
    
    \midrule
    
    MMoE \cite{ma2018modeling}  &   2.67  
    &   5.93  
    &   2.92  
    &   6.35  
    &   3.37  
    &   7.34  
    &   \underline{2.66}
    &   \underline{5.91}
    &   2.84  
    &   6.25  
    &   4.35  
    &   9.02 
    &  23.40 
    &  32.71 
    &  30.34 
    &  42.70 
    &  36.11 
    &  52.24 
    &  15.78 
    &  25.25 
    &  18.46 
    &  27.87 
    &  19.12 
    &  29.83 \\
    
    PLE \cite{tang2020progressive} &   2.51  
    &   5.51  
    &   2.78  
    &   6.12  
    &   3.31  
    &   7.10  
    &   2.57  
    &   5.64  
    &   2.73  
    &   5.91  
    &   4.26  
    &   8.74  
    &  23.79 
    &  34.01 
    &  31.98 
    &  44.29 
    &  41.02 
    &  56.44 
    &  16.28 
    &  24.55 
    &  17.57 
    &  27.89 
    &  19.75 
    &  29.23\\
    
    \midrule
    
    CoNet \cite{hu2018conet}  &   \underline{2.83}
    &   \underline{6.11}
    &   2.81  
    &   6.11  
    &   3.40  
    &   7.09  
    &   2.51  
    &   5.60  
    &   2.76  
    &   6.18  
    &   4.22  
    &   8.68 
    &  23.11 
    &  33.62 
    &  31.38 
    &  45.51 
    &  37.13 
    &  54.62 
    &  14.53 
    &  23.99 
    &  16.27 
    &  28.47 
    &  18.07 
    &  29.42 \\
    
    MiNet \cite{ouyang2020minet} &   2.74  
    &   5.74  
    &   2.83  
    &   6.19  
    &   3.14  
    &   6.81  
    &   2.49  
    &   5.61  
    &   2.69  
    &   5.96  
    &   3.95  
    &   8.31 
    &  24.55 
    &  \underline{35.45}
    &  \underline{32.55}
    &  \underline{47.14}
    &  \underline{41.51}
    &  \underline{57.54}
    &  14.63 
    &  24.24 
    &  17.27 
    &  28.59 
    &  18.80 
    &  30.38\\

    GA-DTCDR \cite{zhu2020graphical}  &   2.81  
    &   6.03  
    &   \underline{3.00}
    &   \underline{6.44}
    &   3.48  
    &   7.50  
    &   2.47  
    &   5.48  
    &   \underline{2.87}
    &   6.17  
    &   \underline{4.47}
    &   \underline{9.24}
    &  24.15 
    &  34.53 
    &  31.87 
    &  44.01 
    &  40.11 
    &  57.49 
    &  15.84 
    &  25.90 
    &  19.35 
    &  31.75 
    &  22.17 
    &  32.96 \\
    
    \midrule
    
    DML \cite{li2021dual}  &   2.60  
    &   5.64  
    &   2.84  
    &   6.23  
    &   3.19  
    &   6.85  
    &   2.41  
    &   5.36  
    &   \underline{2.87}
    &   6.26  
    &   3.54  
    &   7.41
    &  23.45 
    &  34.63 
    &  32.39 
    &  45.98 
    &  38.51 
    &  55.39 
    &  \underline{16.28}
    &  24.60 
    &  \underline{20.00}
    &  30.88 
    &  \underline{22.62}
    &  \underline{33.52} \\
    
    HeroGraph \cite{cui2020herograph}  &   2.62  
    &   5.68  
    &   2.98  
    &   6.42  
    &   3.33  
    &   7.18  
    &   2.59  
    &   5.74  
    &   2.74  
    &   6.13  
    &   4.25  
    &   8.87 
    &  \underline{24.61}
    &  33.52 
    &  32.43 
    &  43.76 
    &  38.09 
    &  54.44 
    &  15.86 
    &  25.12 
    &  18.05 
    &  30.13 
    &  19.81 
    &  28.81 \\
    
    PTUPCDR \cite{zhu2022personalized}&   2.77  
    &   6.03  
    &   2.89  
    &   6.21  
    &   \underline{3.62}
    &   \underline{7.72}
    &   2.38  
    &   5.35  
    &   2.82  
    &   \underline{6.34}
    &   4.32  
    &   8.88 
    &  23.76 
    &  34.17 
    &  32.26 
    &  45.67 
    &  40.81 
    &  53.97 
    &  16.54 
    &  \underline{26.06}
    &  19.11 
    &  \underline{32.26}
    &  20.82 
    &  32.35 \\
    
    \midrule
    
    NMCDR  & \textbf{  2.97  }
    & \textbf{  6.29  }
    & \textbf{  3.40  }
    & \textbf{  6.96  }
    & \textbf{  4.15  }
    & \textbf{  8.60  }
    & \textbf{  2.80  }
    & \textbf{  6.05  }
    & \textbf{  3.39  }
    & \textbf{  6.97  }
    & \textbf{  5.39  }
    & \textbf{  10.46  }
    &\textbf{  25.37 }
    &\textbf{  36.71 }
    &\textbf{  34.18 }
    & \textbf{ 49.75 }
    &\textbf{  44.19 }
    &\textbf{  61.38 }
    &\textbf{  17.82 }
    & \textbf{ 26.98 }
    &\textbf{  21.40 }
    &\textbf{  33.96 }
    &\textbf{  24.68 }
    & \textbf{ 34.90 }\\
    
    Improvement(\%)  &4.95  & 2.95  & 13.33  & 8.07  & 14.64   & 11.40  & 5.26   & 2.37  & 18.12  & 9.94  & 20.58  & 13.20  
    & 3.09  & 3.55  & 5.01  & 5.54  & 6.45  & 6.67  & 9.46  & 3.53  & 7.00  & 5.27  & 9.11  & 4.12 \\
    \bottomrule
    \end{tabular}
    }}
    \end{table*}

\textit{Single-Domain Recommendation Methods}: (i) \textbf{LR} \cite{richardson2007predicting} is a generalized linear approach which stacks several multi-layer perceptrons (MLPs) to model the user-item interaction. (ii) \textbf{BPR} \cite{rendle2012bpr} is a typical collaborative filtering (CF) based method that measures the relevance between users and items by matrix factorization and optimizes pairwise loss with negative samples. (iii) \textbf{NeuMF} \cite{he2017neural} introduces a novel MF component which replaces the inner dot semantic metric with a neural architecture to learn an arbitrary mapping function. 


\textit{Multi-Task Learning Methods}: (i) \textbf{MMoE} \cite{ma2018modeling} utilizes several domain-shared mixture-of-expert encoders along with domain-specific gating network to optimize each domain-specific downstream task. (ii) \textbf{PLE} \cite{tang2020progressive} designs shared encoder and task-specific encoders explicitly and introduces a progressive routing mechanism to extract and separate deeper domain-related knowledge gradually, 

\textit{Cross-Domain Recommendation Methods}: We first use several typical cross-domain models based on fully overlapping conditions as baselines:
(i) \textbf{CoNet} \cite{hu2018conet} utilizes multi-layer feed-forward networks along with cross connections to enables dual knowledge transfer across domains.
(ii) \textbf{MiNet} \cite{ouyang2020minet} jointly models three types of user interest and contains item-level and interest-level attentions to distill useful information from user historical behaviors.
(iii) \textbf{GA-DTCDR} \cite{zhu2020graphical} models user-item interactions via graph neural networks for every single domain and introduces a pairwise attention-based sharing module to transfer information across domains. Then, we adopt several cross-domain models intending to handle partially overlapped CDR tasks as baselines: 
(iv) \textbf{DML} \cite{li2021dual} develops a novel latent orthogonal mapping strategy by dual metric learning method to preserve user relations between different domains.
(v) \textbf{HeroGraph} \cite{cui2020herograph} introduces a shared global graph collecting users and items from multiple domains and transferring the global information to enhance each local domain recommendation performance.
(vi) \textbf{PTUPCDR} \cite{zhu2022personalized} utilizes pre-trained embedding and a meta network to generate a personalized bridge functions which can transfer the personalized preferences for each user across domains.
\subsubsection{Parameter Settings}
To make a fair experimental comparison, we adopt the same hyper-parameters for all the approaches. \textcolor{black}{Specifically, the embedding dimension $D$ is set as 128, the batch size is set as 512, the learning rate is fixed as 0.0001, and the negative sampling number is fixed as 1 for training and 199 for validation and test.} The Adam optimizer is used to update all parameters. For the specific hyper-parameters used in the comparison baselines, we follow their reported values in the official literature. \textcolor{black}{Additionally, for NMCDR, the number of graph aggregation layers in each component is set as 3 for the intra-to-inter node matching module and 2 for intra node complementing module. Besides, we set $ D_{hge} = 128$, $D_{igm} = 128$, $D_{cgm} = 128$, $D_{ref} = 128$, $\mathcal{K}_{head} = 7$ and $w_{1,2,3,4,5,6,7,8} = 1$}.
For all comparison models, we run each experiment five times and select the best results.


\subsection{Performance Comparisons (RQ1)}
\textcolor{black}{Tables \ref{musicmoviecompare}--\ref{loanfundcompare} report the HR@10 and NDCG@10 evaluation metrics on four multi-target CDR tasks. The best results of each column are highlighted in boldface, while the second-best results are underlined. The performance of all models decreases with the decreasing of the overlapping ratio $\mathcal{K}_{u}$, which makes sense as fewer overlapping users may make knowledge transfer across domains more challenging. \textbf{Our NMCDR achieves average 24.84\% improvements on Amazon datasets and average 3.31\% improvements on MYbank datasets compared with second-best baselines over all overlapping settings.} Besides, we have the following insightful findings:}


\subsubsection{For Single-Domain Recommendation Methods} 
\wujiang{(i) LR with stable generalization ability consistently outperforms CF-based methods (i.e., BPR and NeuMF) suffering from the data-sparsity issue.}
\wujiang{(ii) The multi-task methods and cross-domain methods both embody the superior performance to single-domain methods in most cases with the overlapping ratio range 10\%--90\%. However, their performances drop dramatically under extremely small overlapping ratio (e.g. 0.1\%) and tend to be similar with LR, implying that they cannot effectively collect and transfer the cross-domain knowledge.} 


\subsubsection{For Multi-task Learning Methods} 
\wujiang{(i) In most cases, PLE achieves better performance than MMOE, which indicates that task-shared and task-specific components can avoid harmful parameter interference across tasks.}
\wujiang{(ii) Under the larger overlapping conditions ($\mathcal{K}_{u}=50\%$ or $90\%$), the multi-task methods could obtain comparable performance with cross domain recommendation methods such as CoNet, MiNet and GA-DTCDR.} But such multi-task methods exhibit inferior performance compared with cross domain recommendation methods based on partially overlapping settings, i.e., Herograph and PTUPCDR, since they still rely heavily on overlapping users to transfer knowledge across domains.

\subsubsection{For Cross-Domain Recommendation Methods} 
(i) For cross-domain methods based on fully overlapping conditions, GNN based methods (i.e., GA-DTCDR) consistently perform better than the traditional models (i.e., CoNet and MiNet), which demonstrates the remarkable capacity of GNN to model complex user-item interactions and aggregate beneficial neighboring information. \textcolor{black}{(ii) The performance of such fully overlapped CDR models increases with the increasing of overlapping ratio $\mathcal{K}_{u}$, especially when $\mathcal{K}_{u}$ = 90\%, they show comparable results with partial overlapping models (i.e., DML and Herograph) in most experimental cases and GA-DTCDR even achieves the second-best results for ``Cloth-Sport'' and ``Loan-Fund'' scenarios. (iii) Compared with fully overlapping CDR methods, the partial overlapping CDR models, i.e., Herograph and PTUPCDR, consistently exhibit better performance and could achieve second best results in small overlapping experimental settings, i.e., $\mathcal{K}_{u}$ = 0.1\%--50\%, which indicates that modeling and transferring non-overlapping users across domains is essential to improve recommendation quality in general partial overlapped CDR scenarios.} 
\wujiang{(iv) Though PTUPCDR achieves remarkable success in most cases, the model treats all users equally and does not pay special attention to the majority of data-sparse users. Thus, compared with the proposed NMCDR, it possesses an inferior performance.}

\subsubsection{For Our NMCDR} 
\wujiang{(i) Comparing with other cross-domain baselines, our proposed NMCDR consistently achieves great performance improvements on all four CDR scenarios with all evaluation metrics, especially when $\mathcal{K}_{u}$ gets extremely small, e.g., 0.1\% or 1\%. 
Differing from the other CDR baselines relying heavily on overlapped users to bridge connections of multiple domains and then conduct knowledge transferring,
our well-designed intra-to-inter node matching module could well propagate cross-domain information for both overlapped and non-overlapped users.
Furthermore, by introducing the intra node complementing module, we correct the biased representations for each user, especially for the tail users, which conducts missing information completion. 
(ii) Tables \ref{musicmoviecompare} and \ref{loanfundcompare} with the average interactions of items as (16.27, 30.66) and (204.57, 65.41) show the smaller improvement than Tables  \ref{clothsportcompare}--\ref{phoneeleccompare} with the average interactions of items as (16.98, 21.04) and (10.82, 13.46). The average interactions of items means that the total number of user-item interaction divided by the item numbers (for example, the average interactions of items in Music domain is computed by 713,740/43,858 = 16.27). The higher average interactions of items would ease up the effectiveness of the complemented users’ potential missing interactions provided by our model, leads to the lower improvement in Table \ref{musicmoviecompare} and Table \ref{loanfundcompare}. } 

\subsubsection{Comparisons with different density}
\wujiang{Besides, to verify NMCDR's superior performance in CDR scenarios with different data densities, we further conduct studies by varying the data density $D_{s}$ in $\{10\%, 50\%, 70\%\}$. The experimental results of ``Cloth-Sport'' and ``Loan-Fund'' scenarios are given in Table \ref{cs_lf_DS}. Taking the ``Cloth-Sport'' task as example, $D_{s}=50\%$ indicates that the data densities of ``Cloth'' domain and ``Sport'' domain change from 0.06\% to 0.03\% (computed as 0.06$\%$ * 0.5 = 0.03\%) and 0.02\% to 0.01\% (computed as 0.02$\%$ * 0.5 = 0.01\%), respectively.
The performance of all models decrease with the decreasing of data density, which makes sense as sparser data makes the representation learning and knowledge transferring quite challenging. It is also interesting that the performance improvements of our model against second-best baselines decrease with the decreasing of $D_{s}$. This phenomenon further verifies that too sparse user-item interactions, i.e., $D_{s}$ = 10\% or 50\%, would make all model's (including ours) representation learning procedure quite hard and thus the improvement of our model is less remarkable. Nevertheless, our method consistently outperforms all baselines in all sparsity experimental settings. }

\subsubsection{Model Efficiency} 
In this section, all the comparative models are trained and tested on the same machine, which has a single NVIDIA GeForce A100 with 80GB memory and Intel Core i7-8700K CPU with 64G RAM. Moreover, the number of parameters for typical PLE, MiNet, HeroGraph and NMCDR(ours) is in the same order of magnitude, which is 0.16M, 0.78M, 0.64M and 0.56M, respectively. The training/testing efficiencies of PLE, MiNet, HeroGraph and NMCDR(ours) processing the samples of one batch are 2.96$\times$ $10^{-4}$s/1.84$\times$ $10^{-4}$s, 7.65$\times$ $10^{-4}$s/4.56$\times$ $10^{-4}$s, 6.84$\times$ $10^{-4}$s/4.09$\times$ $10^{-4}$s, and 5.34$\times$ $10^{-4}$s/3.92$\times$ $10^{-4}$s, respectively. In summary, NMCDR could achieve superior performance enhancement in (few) partial overlapping CDR settings while keeping promising time efficiency. 

\subsection{Online A/B Test (RQ1)}
\wujiang{We conduct large-scale online A/B tests on financial partially-overlapping CDR scenarios of MYbank of AntGroup\footnote{https://www.antgroup.com/en}. 
In online serving platform of MYbank, large number of users participate in one or multiple financial domains, such as purchasing funds, mortgage loan or discounting bills. 
Specifically, we choose three popular domains with partially overlapped users, i.e., ``Loan'', ``Fund'' and ``Account'', from MYbank serving platform to conduct the online testing. The average statistics of online traffic logs for 1 day are presented in Table \ref{ABtest-anylis}. 
Our method NMCDR along with three baselines are deployed in the online environment for performance comparison and the overall experimental results from December 1st to December 15th are shown in Table \ref{ABtest-results}. Besides, each model conducts 20$\%$ of the online traffic for a standard A/B testing configuration. The standard CVR metric is utilized as the evaluation metrics. We can observe that NMCDR outperforms all the baselines over all domains with the significant improvement about 6.81$\%$, 4.70$\%$ and 6.58$\%$ in three domains. The result verifies NMCDR's capacity of improving the recommendation performance of multiple domains simultaneously in real online environment.}
\vspace{-1pt}
\begin{table}[h!]
\footnotesize
\setlength{\abovecaptionskip}{0pt}
\setlength{\belowcaptionskip}{5pt}
\centering
\captionsetup{font={footnotesize}}
\caption{Average statistics of online traffic logs for 1 day.}
\label{ABtest-anylis}
\begin{threeparttable} 
\setlength\tabcolsep{3pt}{
{
\begin{tabular}{cc|cc|cc|c}
\toprule
\multicolumn{2}{c|}{\textbf{Dataset}}     & \textbf{Users}       & \textbf{Items}       & \textbf{Ratings}   & \textbf{\#Overlapping}           &  \textbf{Density} \\ \midrule 
\multirow{3}{*} &Loan & 45,263,394  & 48,282 &778,136,734    & \multirow{3}{*}{488,836} & 0.04\% \\
                        & Fund & 801,349  & 1,039  & 479,504         &               & 0.06\% \\ & Account & 4,856,675  & 9,816 & 9,149,842  &                         & 0.02\%  \\  
\bottomrule
\end{tabular}
}}
\begin{tablenotes}    
        \footnotesize            
        \item \#Overlapping denotes the number of overlapped users across domains. 
      \end{tablenotes}
\end{threeparttable}

\end{table}

\begin{table}[h!]
\vspace{-10pt}
\centering
\captionsetup{font={footnotesize}}
\caption{Experimental results of the online A/B testing from 12.1 to 12.15, 2022}
\label{ABtest-results}
\footnotesize
\setlength\tabcolsep{2pt}{
\begin{tabular}{ccccc} 
\toprule
              & \textbf{Loan Domain} & \textbf{Fund~Domain} & \textbf{Account~Domain}   \\ 
\midrule
Control Group & 10.50\%            &  6.12\%           & 1.89\%                           \\ 

MMOE~Group    & 12.14\%            &   6.45\%          & 2.11\%                           \\ 
PLE~Group    & 12.57\%            &    6.69\%         & 2.27\%                           \\ 
DML~Group     & \underline{12.93\%}            & \underline{6.81\%}             & \underline{2.43\%}                             \\ 
\midrule
NMCDR~Group   & \textbf{13.81\%}            & \textbf{7.13\%}            &\textbf{2.59\%}                              \\
Improvement  & 6.81\%            &  4.70\%            & 6.58\%                             \\
\bottomrule
\end{tabular}}
\end{table}
\vspace{-10pt}

\subsection{Model Analysis (RQ2)}
\subsubsection{Impact of Different Model Components}
To verify the contribution of each key component of NMCDR, we conduct an ablation study by comparing with several variants: (i) $w/o$-Igm: we remove the intra node matching component for conducting intra knowledge fusion for both head and tail users in every domain. (ii) $w/o$-Cgm: we remove the inter node matching component for conducting inter domain knowledge fusion and transferring across domains. (iii) $w/o$-Inc: we remove the intra node complementing module for correcting the biased user representations in each domain. (iv) $w/o$-Sup: we remove the multiple supervisory signals into each key module for guiding knowledge fusion and transfer procedure. We conduct the ablation experiments with overlapping ratio $\mathcal{K}_{u} = 50\%$ and report the results in Table \ref{abl_components}. Based on Table \ref{abl_components}, we draw the following observations:
\wujiang{(a) It is critical to perform intra knowledge fusion for both head and tail users before conducting subsequent cross domain knowledge transferring, especially for tail users, since their information deriving from sparsely observed interactions may be biased and harmful to other domains, which hurts the performance as shown in $w/o$-Igm column. }
(b) When removing the inter node matching component, our model cannot collect and transfer knowledge for both overlapping and non-overlapping users across domains, which hurts the performance significantly. (c) Without the intra node complementing module, the under-represented user embeddings would be used to conduct ranking recommendation tasks and thus impair performance. (d) Without the
multiple supervisory signals into each key module, the performance also drops obviously, which indicates that effective supervision signals are essential to guide the learning process of each module and result in satisfying results. (e) Overall, Cgm provides the largest contributes for our method. Besides, the multiple supervisory signals (Sup) into each key module for guiding knowledge fusion and transfer procedure brings slightly larger contribution than Igm and Inc.

\begin{table}[h!]
\captionsetup{font={footnotesize}}
\caption{Experimental results (\%) with different model variants. $w/o$ denotes the model without the corresponding component variant.}
\label{abl_components}
\setlength\tabcolsep{3pt}{
\begin{tabular}{cccccccc}
\toprule
\multicolumn{1}{c}{\multirow{2}{*}{Scenarios}} & \multirow{2}{*}{Metrics} & \multicolumn{4}{c}{Model variants}                               & \multirow{2}{*}{Ours}                                                                       \\ \cline{3-6} 
\multicolumn{1}{c}{}                           &                          & \multicolumn{1}{c}{$w/o$-Igm} & \multicolumn{1}{c}{$w/o$-Cgm} & \multicolumn{1}{c}{$w/o$-Inc} & \multicolumn{1}{c}{$w/o$-Sup} \\ \midrule
\multirow{2}{*}{Music}                    
& NDCG@10                        & \multicolumn{1}{c}{10.28}         & \multicolumn{1}{c}{9.30}         & \multicolumn{1}{c}{10.90}         & \multicolumn{1}{c}{9.78}    &\multicolumn{1}{c}{\bf 11.26}          \\ 
& HR@10                          & \multicolumn{1}{c}{19.28}         & \multicolumn{1}{c}{18.78}         & \multicolumn{1}{c}{20.89}         & \multicolumn{1}{c}{19.16}    &\multicolumn{1}{c}{\bf 21.58}          \\  \cline{2-7}
\multirow{2}{*}{Movie}                   
& NDCG@10                         & \multicolumn{1}{c}{32.84}         & \multicolumn{1}{c}{31.96}         & \multicolumn{1}{c}{33.60}         & \multicolumn{1}{c}{32.60}    &\multicolumn{1}{c}{\bf 33.96}          \\ 
& HR@10                           & \multicolumn{1}{c}{48.73}         & \multicolumn{1}{c}{48.01}         & \multicolumn{1}{c}{50.48}         & \multicolumn{1}{c}{48.93}    &\multicolumn{1}{c}{\bf 51.13}          \\ \midrule
\multirow{2}{*}{Cloth}                    
& NDCG@10                        & \multicolumn{1}{c}{9.14}         & \multicolumn{1}{c}{7.35}         & \multicolumn{1}{c}{8.95}         & \multicolumn{1}{c}{8.38}    &\multicolumn{1}{c}{\bf 9.26}          \\ 
& HR@10                          & \multicolumn{1}{c}{17.99}         & \multicolumn{1}{c}{15.14}         & \multicolumn{1}{c}{17.65}         & \multicolumn{1}{c}{17.59}    &\multicolumn{1}{c}{\bf 18.33}          \\ \cline{2-7}
\multirow{2}{*}{Sport}                   
& NDCG@10                         & \multicolumn{1}{c}{14.75}         & \multicolumn{1}{c}{13.02}         & \multicolumn{1}{c}{14.60}         & \multicolumn{1}{c}{13.98}    &\multicolumn{1}{c}{\bf 14.91}          \\ 
& HR@10                           & \multicolumn{1}{c}{26.94}         & \multicolumn{1}{c}{24.35}         & \multicolumn{1}{c}{26.86}         & \multicolumn{1}{c}{27.04}    &\multicolumn{1}{c}{\bf 27.54}          \\ \midrule
\multirow{2}{*}{Phone}                    
& NDCG@10                        & \multicolumn{1}{c}{16.50}         & \multicolumn{1}{c}{14.42}         & \multicolumn{1}{c}{17.05}         & \multicolumn{1}{c}{17.09}    &\multicolumn{1}{c}{\bf 17.44}          \\ 
& HR@10                          & \multicolumn{1}{c}{29.47}         & \multicolumn{1}{c}{25.37}         & \multicolumn{1}{c}{29.70}         & \multicolumn{1}{c}{29.82}    &\multicolumn{1}{c}{\bf 30.87}          \\ \cline{2-7}
\multirow{2}{*}{Elec}                   
& NDCG@10                         & \multicolumn{1}{c}{23.75}         & \multicolumn{1}{c}{20.82}         & \multicolumn{1}{c}{24.10}         & \multicolumn{1}{c}{24.13}    &\multicolumn{1}{c}{\bf 24.45}          \\ 
& HR@10                           & \multicolumn{1}{c}{37.95}         & \multicolumn{1}{c}{33.87}         & \multicolumn{1}{c}{38.26}         & \multicolumn{1}{c}{38.43}    &\multicolumn{1}{c}{\bf 39.49}          \\ \midrule
\multirow{2}{*}{Loan}                    
& NDCG@10                        & \multicolumn{1}{c}{49.69}         & \multicolumn{1}{c}{49.40}         & \multicolumn{1}{c}{49.76}         & \multicolumn{1}{c}{49.67}    &\multicolumn{1}{c}{\bf 49.89}          \\ 
& HR@10                          & \multicolumn{1}{c}{69.83}         & \multicolumn{1}{c}{69.32}         & \multicolumn{1}{c}{69.89}         & \multicolumn{1}{c}{69.79}    &\multicolumn{1}{c}{\bf 69.98}          \\  \cline{2-7}
\multirow{2}{*}{Fund}                   
& NDCG@10                         & \multicolumn{1}{c}{34.84}         & \multicolumn{1}{c}{34.77}         & \multicolumn{1}{c}{35.10}         & \multicolumn{1}{c}{34.90}    &\multicolumn{1}{c}{\bf 35.24}          \\  
& HR@10                           & \multicolumn{1}{c}{54.84}         & \multicolumn{1}{c}{54.35}         & \multicolumn{1}{c}{54.91}         & \multicolumn{1}{c}{54.80}    &\multicolumn{1}{c}{\bf 55.03}          \\ \bottomrule
\end{tabular}}
\end{table}
\vspace{0.1pt}
\subsection{Hyperparameter Analysis (RQ3)}
\subsubsection{Number of Matching Neighbors}
\begin{figure}[h!]
\centering{
\includegraphics[width=0.4\textwidth]{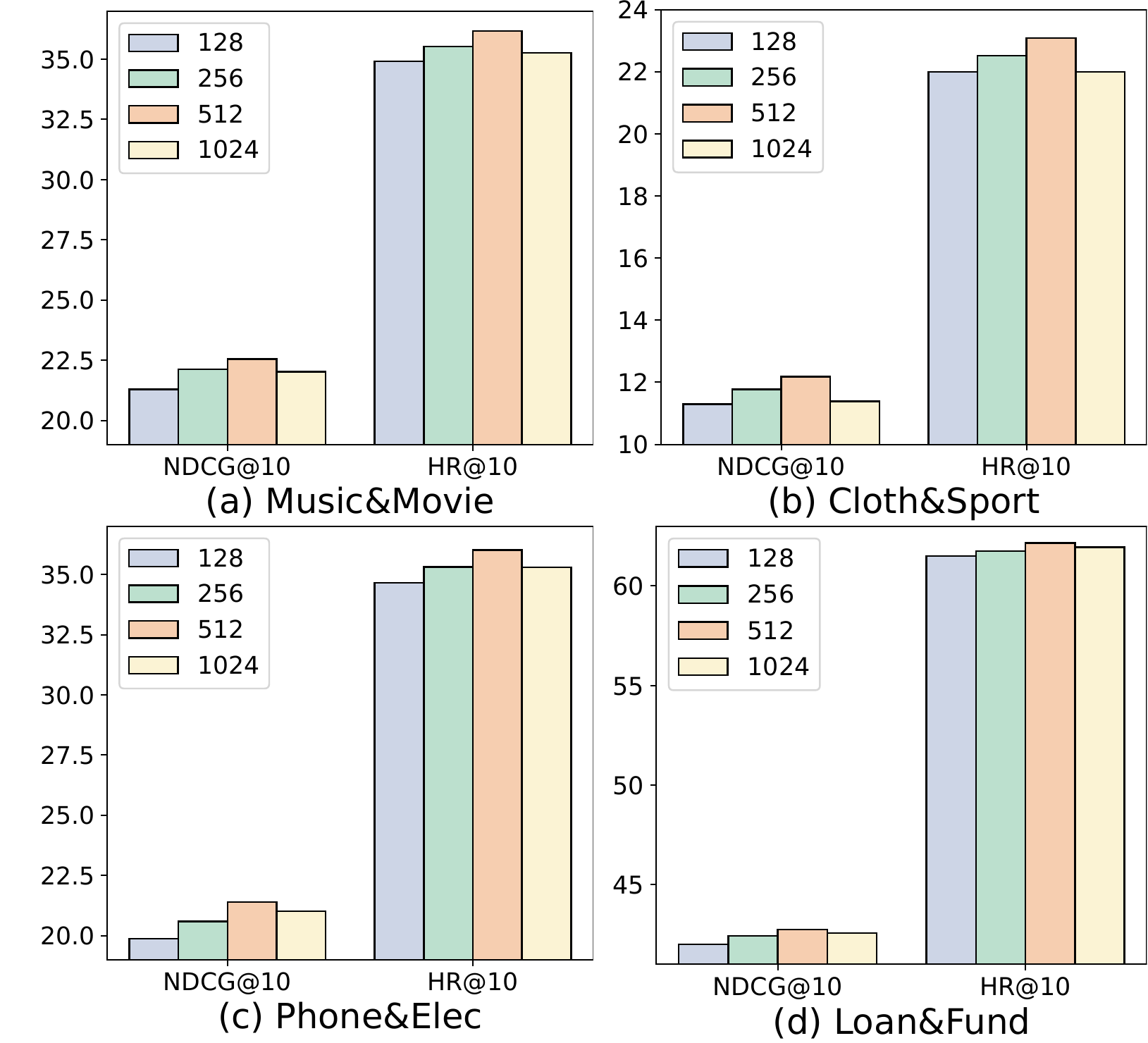}}
\captionsetup{font={footnotesize}}
\caption{Impact of number of matching neighbors. }
\label{fig3}
\end{figure}

To explore the impact of the number of the neighborhood for intra and inter node matching, we conduct ablation experiments by varying the number of matching neighbors from 128 to 1024. The average evaluation results (i.e., NDCG@10 and HR@10) for each dataset are shown in Fig. \ref{fig3} and we can observe that as the number of matching neighbors increases, the recommendation performance initially rises steadily and then descend when the matching neighbors reach 1024. This phenomenon indicates that too small matching neighbors would provide limited transferred information, while too many matching neighbors may introduce interference noise and impair the model performance. In practice, we set 512 to balance the training efficiency and model performance.
\subsubsection{Threshold of Head/Tail User Discrimination}
\begin{figure}[h!]
\centering{
\includegraphics[width=0.40\textwidth]{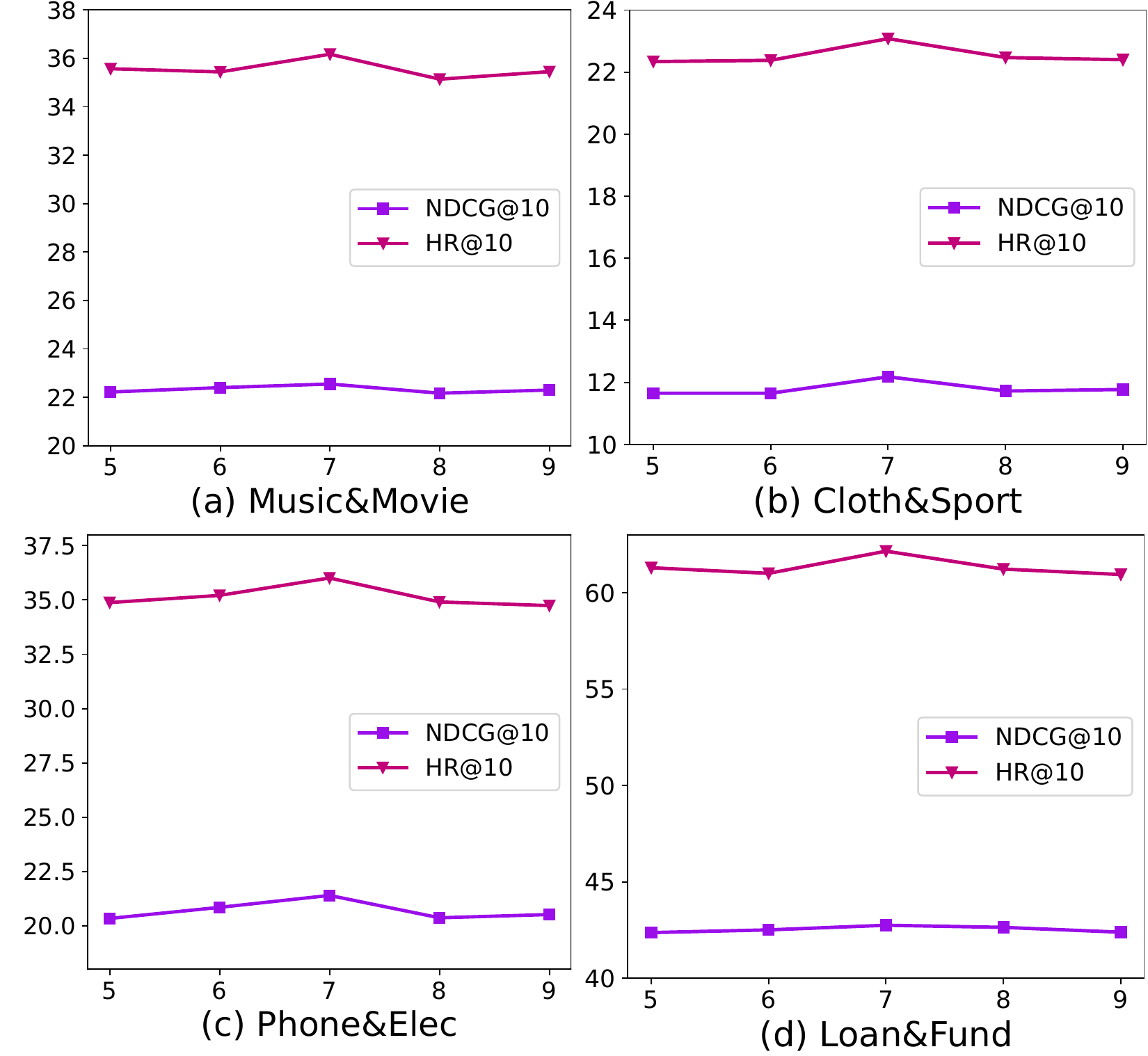}}
\captionsetup{font={footnotesize}}
\caption{Impact of threshold of head/tail user discrimination. }
\label{neighbornumber}
\end{figure}

In this part, we explore the impact of head/tail user discrimination threshold $\mathcal{K}_{head}$ on model performance. If the historical interactions of a user is greater than $\mathcal{K}_{head}$, then he/she is regarded as a head user. Otherwise he/she would be treated as a tail user.  The experimental results are shown in Fig. \ref{neighbornumber}. Firstly, the average performance gains of all tasks slightly rise then descend with the increase of $\mathcal{K}_{head}$. The small variations of model performance indicate the robustness of NMCDR. Besides, the variation tendency of model performance is similar for different datasets, which may be caused by the data pre-processing procedure as we remove the user with less than 5 interactions for each dataset.

\vspace{-10pt}
\vspace{-10pt}

\subsection{Visualization Analysis}
\begin{figure}[h!]
\centering{
\includegraphics[width=0.49\textwidth]{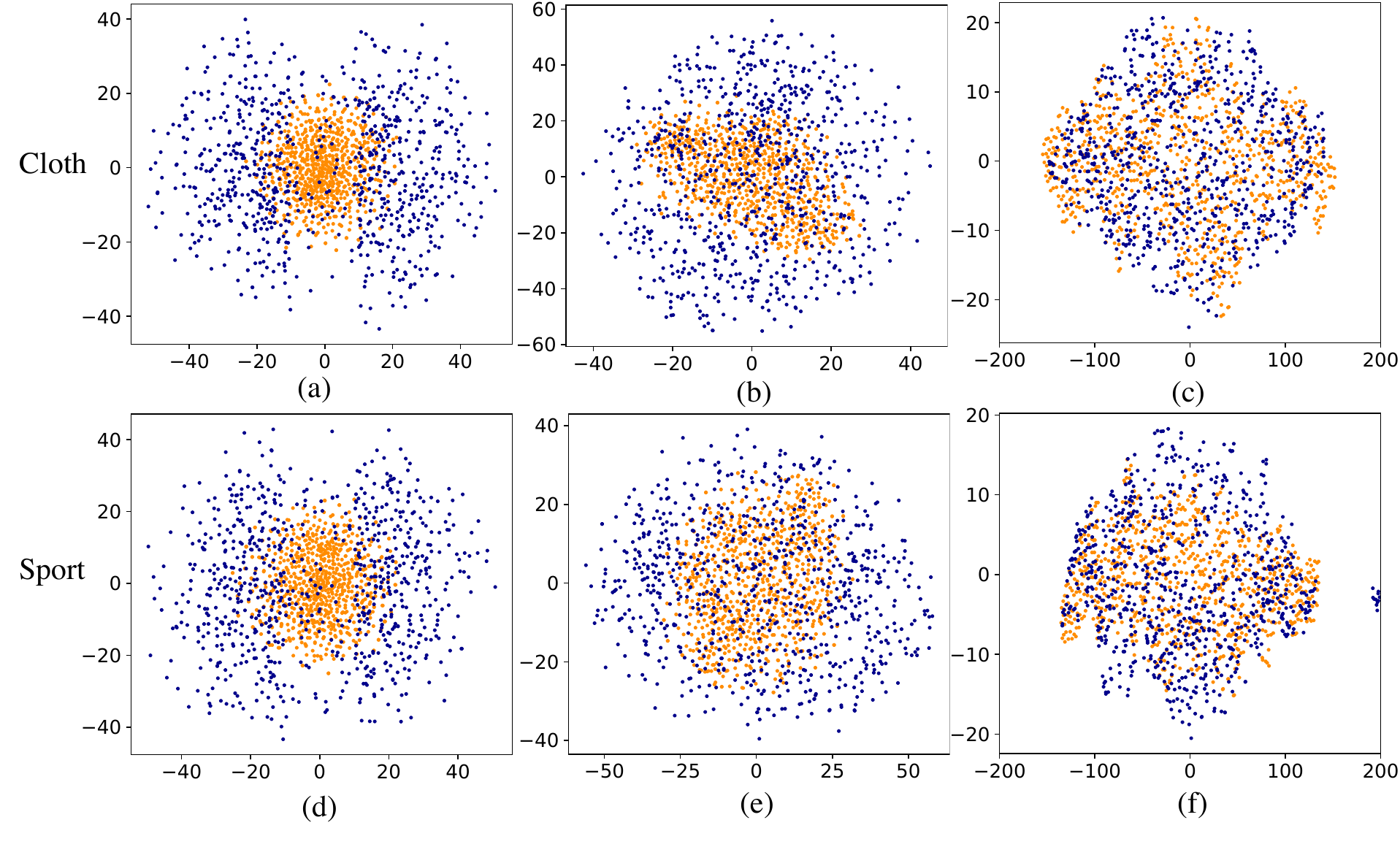}}
\captionsetup{font={footnotesize}}
\caption{The visualization of learned user representations for evaluating the effectiveness of NMCDR's each key component.}
\label{visual_tsne}
\end{figure}

Besides the quantitative evaluation, we also provide intuitive user embeddings and try to visualize the effect of each key component of NMCDR. Fig. \ref{visual_tsne} shows the t-SNE visualization of the head (yellow dots) and tail (blue dots) user embeddings on Amazon ``Cloth-Sport'' scenario with overlapping ratio $\mathcal{K}_{u} = 50\%$. In Fig. \ref{visual_tsne}, the first column (Figs. \ref{visual_tsne}(a), (d)), second column (Figs. \ref{visual_tsne}(b), (e)) and third column (Figs. \ref{visual_tsne}(c), (f)) indicate the obtained user embeddings after being processed by initial graph encoder layer, intra-to-inter node matching module and intra node complementing module respectively. From it, we have the following observations: (i) After a typical heterogeneous graph encoder, the head and tail user embeddings for both ``Cloth'' and ``Sport'' domains show clear distinction as shown in Figs. \ref{visual_tsne}(a) and (d), but the tail user embedding may be under-represented based on the observed sparse neighboring nodes and such issue is often neglected in previous work. (ii) As shown in Figs. \ref{visual_tsne}(b) and (e), the head and tail user embeddings tend to align by conducting fully connected intra and inter knowledge transferring. (iii) After intra node complementing module as shown in Figs. \ref{visual_tsne}(c) and (f), the embedding distributions of tail users do exhibit quite similar to that of head users by complementing potential missing interaction information, which is essential to get superior recommendation performance and in line with our motivation.

\vspace{-10pt}

\section{Related work}
\noindent \textcolor{black}{\textbf{Multi-Target Cross Domain Recommendation} is an effective method to improve the recommendation performance in multiple domains simultaneously and to alleviate the long-standing data sparsity and cold-start problem in recommender systems.}
Generally, the existing works of multi-target CDR can be roughly divided into the following two groups: cross-domain models based on fully overlapping settings and cross-domain models intending to handle partially overlapped CDR tasks. To transfer knowledge across domains based on fully overlapping settings, several excellent works focusing on feature combination \cite{zhu2020deep,zhang2018cross,zhao2019cross} or bi-directional transfer mapping strategies \cite{hu2018conet,zhu2020graphical,ouyang2020minet,salah2021towards} have been proposed and achieved promising results. Especially, PPGN \cite{zhao2019cross} combines the dual-domain features into the graph neural networks to learn the cross-domain information, while CoNet \cite{hu2018conet}, GA-DTCDR \cite{zhu2020graphical} and MiNet \cite{ouyang2020minet} mainly focus on designing the mapping functions to fuse and transfer useful information across domains. As the above learning frameworks primarily assume the existence of fully overlapped users or items across domains, leading them incapable of handling partially overlapped CDR scenarios. To alleviate this issue and develop models for partially overlapped CDR settings, several recent efforts \cite{zhu2019dtcdr, li2020ddtcdr,li2021dual, cui2020herograph} try to introduce graphic neural networks to get both overlapped and non-overlapped user embeddings by collecting user-item interactions. Additionally, PTUPCDR \cite{zhu2022personalized}  designs a meta network to generate personalized bridge functions for each user. SA-VAE \cite{salah2021towards} and VDEA \cite{liu2022exploiting} utilizes variational auto-encoder (VAE) framework to exploit user domain-invariant embedding across different domains. However, such partially overlapped models treat all users equally and do not pay special attention to the majority of data-sparse users and resulting in inferior knowledge fusing and transferring effectiveness.

\noindent \textcolor{black}{\textbf{Neural Graph Matching} 
intends to discover the node level or graph level similarity between two given graphs \cite{caetano2009learning, yan2005substructure,dijkman2009graph}.} Before GNNs-based methods, traditional graph matching approaches usually measure graph similarity based on heuristic rules, i.e., minimal graph edit distance \cite{willett1998chemical,raymond2002rascal}, or graph kernel based matching methods, i.e., random walks inside graphs \cite{vishwanathan2010graph,kashima2003marginalized} and graph sub-structures \cite{shervashidze2009fast,shervashidze2009efficient}. In recent years, GNNs-based graph matching methods are frequently proposed and achieved great success. Li et al. \cite{li2019graph} consider computing the similarity of two given graphs by a carefully designed cross-graph attention-based matching mechanism. Xu et al. \cite{xu2019cross} formulate the KB-alignment task as a graph matching problem and models the local matching information through a graph-attention based solution. Soldan et al. \cite{soldan2021vlg} introduces a Video-Language graph matching network and utilize the mutual exchange of information to enhance the multi-modal representation for video grounding task. \textcolor{black}{Recently, Su et al. \cite{su2021neural} proposes a neural graph matching CF-based model to capture attribute interactions for recommendation system. However, such CF-based graph matching framework cannot be directly utilized in CDR scenarios when encountering data sparsity issues.}
\vspace{1pt}

\section{Conclusion}
In this paper, to develop a simple-yet-effective multi-target CDR framework for the more general CDR settings with only partially overlapped users or items, we propose a novel node matching based framework, namely NMCDR. The developed model mainly contains two modules, i.e., intra-to-inter node matching and intra node complementing module. The intra-to-inter node matching module could effectively fuse and transfer the knowledge within-domain as well as cross-domain for all users, especially for the non-overlapping users, without relying heavily on overlapping users. Additionally, intra node complementing module complements the potential missing information for each user to correct his/her biased representation for ranking recommendation tasks, especially for the tail users with observed sparse interactions. To our knowledge, this paper is the first work to correct the potential interactions bias in multi-target CDR scenarios. Extensive experiments
demonstrate the remarkable effectiveness of the proposed approach in kinds of evaluation metrics and elaborate
ablation studies present the contribution of each module to the final performance gain.  



\newpage

\bibliographystyle{IEEEtran}
\bibliography{IEEEabrv}

\end{document}